\documentclass[twocolumn]{aastex63}

\received{October 18, 2021}

\shorttitle{From naked spheroids to disky galaxies}
\shortauthors{Costantin et al.}

\usepackage[]{xcolor}
\usepackage{comment}

\defcitealias{Costantin2021a}{Paper I}

\begin{document}

\title{From naked spheroids to disky galaxies: how do massive disk galaxies shape their morphology?}

\correspondingauthor{Luca Costantin}
\email{lcostantin@cab.inta-csic.es}

\author[0000-0001-6820-0015]{Luca Costantin}
\affiliation{Centro de Astrobiolog\'ia (CSIC-INTA), Ctra de Ajalvir km 4, Torrej\'on de Ardoz, 28850, Madrid, Spain}
\affiliation{INAF - Osservatorio Astronomico di Brera, Via Brera 28, 20121, Milano, Italy}

\author[0000-0003-4528-5639]{Pablo G.~P\'erez-Gonz\'alez}
\affiliation{Centro de Astrobiolog\'ia (CSIC-INTA), Ctra de Ajalvir km 4, Torrej\'on de Ardoz, 28850, Madrid, Spain}
\affiliation{Honorary professor, Departamento de F\'isica de la Tierra y Astrof\'isica, Facultad de CC. F\'isicas, Universidad Complutense de Madrid, 28040 Madrid, Spain}

\author[0000-0002-8766-2597]{Jairo M\'endez-Abreu}
\affiliation{Instituto de Astrof\'isica de Canarias, 38200, La Laguna, Tenerife, Spain}
\affiliation{Departamento de Astrof\'isica, Universidad de La Laguna, 38205, La Laguna, Tenerife, Spain}

\author[0000-0002-1416-8483]{Marc Huertas-Company}
\affiliation{Instituto de Astrof\'isica de Canarias, 38200, La Laguna, Tenerife, Spain}
\affiliation{Departamento de Astrof\'isica, Universidad de La Laguna, 38205, La Laguna, Tenerife, Spain}
\affiliation{LERMA, Observatoire de Paris, CNRS, PSL, Universit\'e de Paris, France}

\author[0000-0002-4140-0428]{Bel\'en Alcalde Pampliega}
\affiliation{European Southern Observatory (ESO), Alonso de C\'ordova 3107, Vitacura, Casilla 19001, Santiago de Chile, Chile}

\author[0000-0002-3935-9235]{Marc Balcells}
\affiliation{Isaac Newton Group of Telescopes, Apartado 321, 38700, Santa Cruz de La Palma, Islas Canarias, Spain}
\affiliation{Instituto de Astrof\'isica de Canarias, 38200, La Laguna, Tenerife, Spain}
\affiliation{Departamento de Astrof\'isica, Universidad de La Laguna, 38205, La Laguna, Tenerife, Spain}

\author[0000-0001-6813-875X]{Guillermo Barro}
\affiliation{Department of Physics, University of the Pacific, 3601 Pacific Ave., Stockton, CA 95211, USA}

\author[0000-0002-8680-248X]{Daniel Ceverino}
\affiliation{Universidad Autonoma de Madrid, Ciudad Universitaria de Cantoblanco, 28049, Madrid, Spain}
\affiliation{CIAFF, Facultad de Ciencias, Universidad Autonoma de Madrid, 28049 Madrid, Spain}

\author[0000-0001-7399-2854]{Paola Dimauro}
\affiliation{Observat\'orio Nacional, Rua General Jos\'e Cristino, 77, S\~ao Crist\'ov\~ao, 20921-400, Rio de Janeiro, Brazil}

\author[0000-0002-9013-1316]{Helena Dom\'inguez S\'anchez}
\affiliation{Institute of Space Sciences (ICE, CSIC), Campus UAB, Carrer de Magrans, 08193, Barcelona, Spain}
\affiliation{Institut d'Estudis Espacials de Catalunya (IEEC), 08034, Barcelona, Spain}

\author[0000-0001-6426-3844]{N\'estor Espino-Briones}
\affiliation{Instituto Nacional de Astrof\'isica, \'Optica y Electr\'onica, Luis E. Erro No.~1, Tonantzintla, 72840 Puebla, M\'exico}

\author[0000-0002-6610-2048]{Anton M. Koekemoer}
\affiliation{Space Telescope Science Institute, 3700 San Martin Dr., Baltimore, MD 21218, USA}


\begin{abstract}
We investigate the assembly history of massive disk galaxies and describe how they
shape their morphology through cosmic time.
Using SHARDS and HST data, we modeled the surface brightness distribution of 91 massive galaxies at
redshift $0.14<z\leq 1$ in the wavelength range
$0.5-1.6$~$\mu$m, deriving the uncontaminated spectral energy distributions of their bulges and disks separately.
This spectrophotometric decomposition allows us to compare the stellar populations properties
of each component in individual galaxies.
We find that the majority of massive galaxies ($\sim85\%$) builds inside-out,
growing their extended stellar disk around the central spheroid.
Some bulges and disks could start forming at similar epochs, 
but these bulges grow more rapidly than their disks,
assembling 80\% of their mass in $\sim0.7$~Gyr and $\sim3.5$~Gyr, respectively.
Moreover, we infer that both older bulges and older disks are more massive and compact than younger stellar structures.
In particular, we find that bulges display a bimodal distribution of mass-weighted ages,
i.e., they form in two waves.
In contrast, our analysis of the disk components indicates that they form
at $z\sim1$ for both first and second-wave bulges. This translates to first-wave bulges taking longer
in acquiring a stellar disk ($5.2$~Gyr) compared to second-wave less-compact spheroids ($0.7$~Gyr).
We do not find distinct properties (e.g., mass, star formation timescale, and mass surface density) for the disks in both types of galaxies.
We conclude that the bulge mass and compactness mainly regulate the timing of the stellar disk growth, 
driving the morphological evolution of massive disk galaxies.
\end{abstract}




\section{Introduction \label{sec:section_1}}

The morphological classification of galaxies represents the first attempt to
understand the origins of the variety of observed galaxies in the universe \citep{Hubble1926}.
However, a crucial but still unresolved controversy 
is the origin of the Hubble sequence. When galaxies shape their morphology? What
drives their evolution? How do the spheroidal and disk components
of present-day galaxies form?

Traditionally, the stellar disk is proposed to form through the 
collapse of the gas in a rotating dark matter halo \citep{Fall1980}.
Although the nature of the processes involved is dissipative,
the gas conserves its mass and high angular momentum  
in the absence of external influences \citep{Dalcanton1997, Mo1998}.
Consequently, it enhances a differential star formation 
in the galaxy, since the central region reaches a sufficient 
gas surface mass density to form stars earlier or in larger amounts 
and with higher efficiency than the outer part \citep{Brook2006}.
The above description is far from simple within the hierarchical assembly of structures
in a cold dark matter universe. 
Galaxy mergers usually destroy (or thicken) the stellar disk \citep{Steinmetz2002},
but the stellar and gas material could survive to re-form a disk in the merger 
remnant \citep{Hopkins2009b, Clauwens2018}.

Multiple possibilities are proposed for the formation of the central bulge component.
Accordingly, bulges are usually classified as classical or disk-like
depending on their main channel of evolution \citep{Athanassoula2005}.
Classical bulges could arise from a violent and dissipative collapse of 
protogalaxies \citep{Eggen1962, Larson1976},
from accumulation and rearrangement of stars in mergers events \citep{Cole2000, Hopkins2009a},
from massive clumps coalescence \citep{Noguchi1999, Bournaud2007}, 
and/or a gas-compaction phase triggered by violent disk instabilities which lead 
to efficient spheroidal growth in high-redshift galaxies \citep{Dekel2014, Ceverino2015, Zolotov2015}.
On the other hand, the slow and prolonged rearrangement of disk material due 
to secular evolutionary processes (i.e., the evolution of a bar component, instabilities due to spiral patterns, etc.) 
could build-up central bulges with disk-like
properties \citep{Kormendy2004, Kormendy2016}.
From an observational standpoint, the challenge is to reconstruct
the formation pathways of galaxies having access to the observed properties
of their bulges and disks \citep{MendezAbreu2010, MendezAbreu2014, 
Morelli2015, Morelli2016, Costantin2017, Costantin2018a, Costantin2018b,
deLorenzoCacered2019a, deLorenzoCacered2019b, Gadotti2020, Gao2020}.

In the last years, there has been a large effort
to study the stellar population properties of high-redshift galaxies
with the aim of unveiling the main processes which drove 
their evolution \citep[e.g.,][]{Belli2019, EstradaCarpenter2019, Tacchella2021}.
These studies suggest that at high redhift galaxies form on shorter timescales
with respect to those formed at later cosmic times.
Moreover, the formation redshift of these galaxies
seems to depend on their stellar mass \citep{Heavens2004, PerezGonzalez2008, Morishita2019, Carnall2019}
and mass surface density \citep{EstradaCarpenter2020, Suess2021}.

The main problem of studying the galaxy integrated properties is
that the complexity of its formation and evolution is averaged out.
But, until recently, few works have focused their attention
in studying the separate evolution of the different morphological components 
beyond the local universe \citep[e.g.,][]{DominguezPalmero2008, 
DominguezPalmero2009, Bruce2014, MargalefBentabol2016, MargalefBentabol2018, Dimauro2018, Mancini2019}.
Thus, in this work we show how the accreted mass fraction in bulges and disks can actually
unveil fundamental hints about the balance between the 
different pathways of galaxy formation.

\begin{figure*}[t!]
\centering
\includegraphics[scale=0.63]{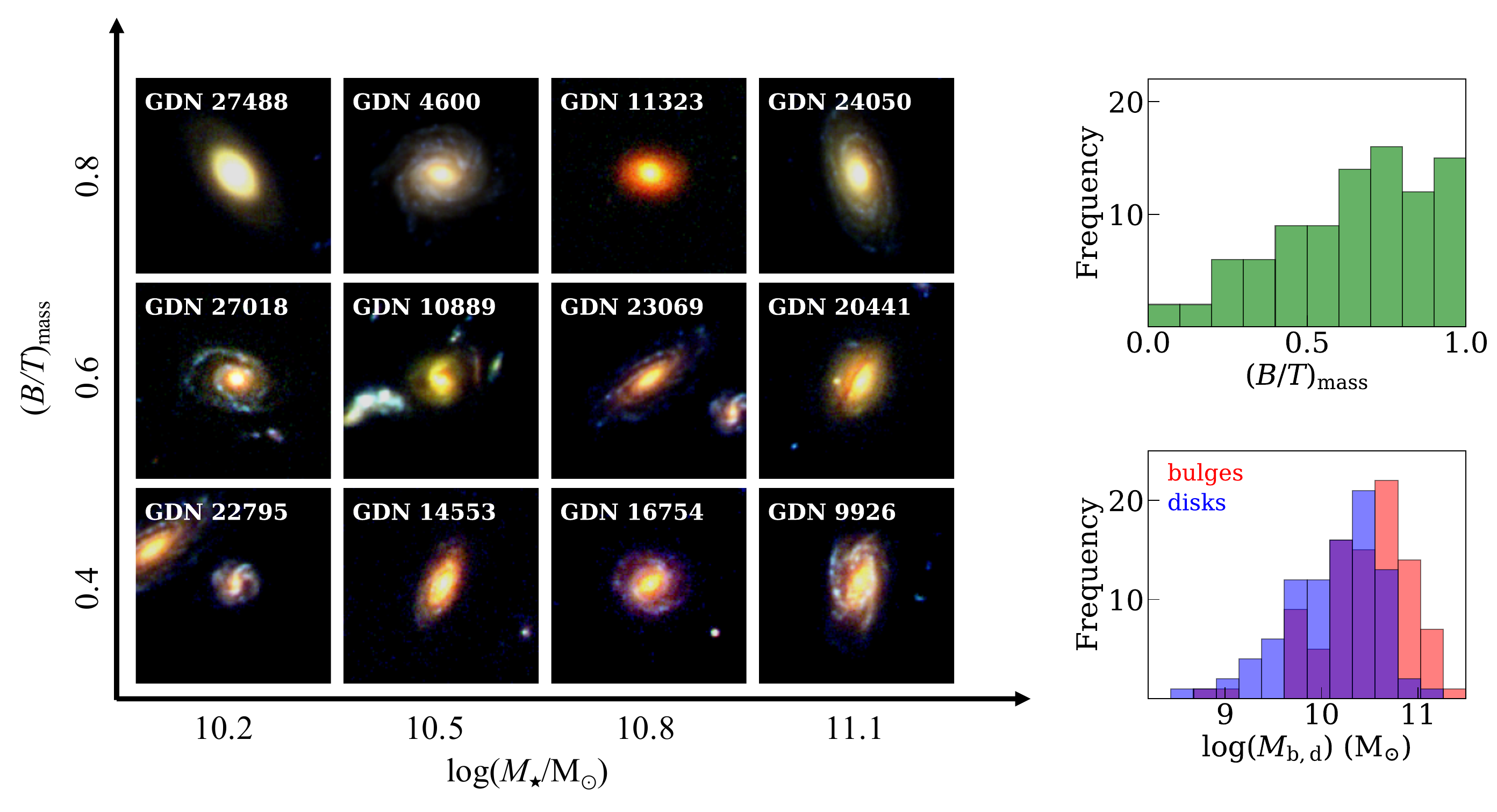}
\caption{Left panel: Visual examples from the HST imaging \citep{Koekemoer2011} 
of the diversity of morphologies of our galaxies as a function of 
their stellar mass and bulge-over-total mass ratio $(B/T)_{\rm mass}$.
Right panels: Bulge-over-total mass ratio (green) and mass distributions of bulges (red)
and disks (blue).
\label{fig:figure_1}}
\end{figure*}

In this paper, we study the interplay of bulge and disk properties across time
exploiting the Survey for High-z Absorption Red and Dead Sources 
\citep[SHARDS;][]{PerezGonzalez2013} data set.
SHARDS is a state-of-the-art multi-filter imaging survey,
which provides ultra-deep ($m < 26.5$ AB mag) photometry 
in 25 filters covering the wavelengths range $0.50-0.95$~$\mu$m with subarsec seeing.
SHARDS data allow us to smoothly sample 
the spectral energy distribution (SED) of galaxies with spectral resolution $R\sim50$
and, given the seeing upper limits imposed in the Gran Telescopio Canarias queue-mode data acquisition, 
separate the light of their individual bulge and disk components.
In \citet[][hereafter \citetalias{Costantin2021a}]{Costantin2021a} we presented the spectrophotometric decoupling of 
a sample of massive galaxies up to redshift $z=1$ and presented their bulge properties. 
In particular, we found a bimodal distribution of bulge ages, with a fraction of them being formed
in the early universe ($z\sim6$) and having high mass surface densities;
a second wave of bulges, dominant in number, 
evolved more slowly, forming most of their stars $\sim5$~Gyr later.
In this second paper of a series, we aim at studying the interplay
between the bulge and the disk properties through time,
in order to describe their relative importance in building up massive disk galaxies.
For that purpose, we take the same sample presented in \citetalias{Costantin2021a}, 
concentrating in the galaxies with disks (i.e., leaving aside pure spheroids), 
and study their stellar population properties.

The paper is organized as follows. We describe the data set and summarize
the spectro-photometric decoupling procedure and the stellar population analysis in Sect.~\ref{sec:section_2}.
We present and discuss our results in Sect.~\ref{sec:section_3} 
and \ref{sec:section_4}, respectively. We provide our conclusions in Sect.~\ref{sec:section_5}.
Throughout the paper we assume a flat cosmology with $\Omega_{\rm m} = 0.3$,
$\Omega_{\rm \lambda} = 0.7$, a Hubble constant $H_0 = 70$ km s$^{-1}$ Mpc$^{-1}$, and 
a \citet{Chabrier2003} initial mass function ($0.1 < M/M_{\odot} < 100$).


\section{Data \label{sec:section_2}}

In this Section we present the data set and the properties of the sample of 
galaxies analyzed in this work (Sect.~\ref{sec:section_2_1}). 
The characterization of the bulge and disk
physical properties is fully described in \citetalias{Costantin2021a}.
For completeness, we briefly summarize the main steps of the 
spectrophotometric decomposition (Sect.~\ref{sec:section_2_2}), 
as well as the analysis of the stellar population properties (Sect.~\ref{sec:section_2_3}).

\subsection{Sample of Galaxies \label{sec:section_2_1}}

We combine the spectral resolution of the SHARDS observations
with the high spatial resolution of the Hubble Space Telescope (HST) 
Advanced Camera for Surveys (ACS) and Wide Field Camera 3 (WFC3) images.
In particular, we use seven filters for HST images
from the optical to the near-infrared wavelength range $0.475-1.600$~$\mu$m
and the 25 filters of SHARDS in the optical wavelength range $0.500-0.941$~$\mu$m
\citep[see][for all details]{Grogin2011, Koekemoer2011, PerezGonzalez2013, Barro2019}.
To provide a more robust constraint on the stellar mass,
we complement this data set with the $K$-band information at $\sim 2.1$~$\mu$m
provided by the Canada-France-Hawaii Telescope WIRCam data \citep{Hsu2019}.

\begin{figure*}[t!]
\centering
\includegraphics[scale=0.36, trim=0cm 0cm 0cm 0cm , clip=true]{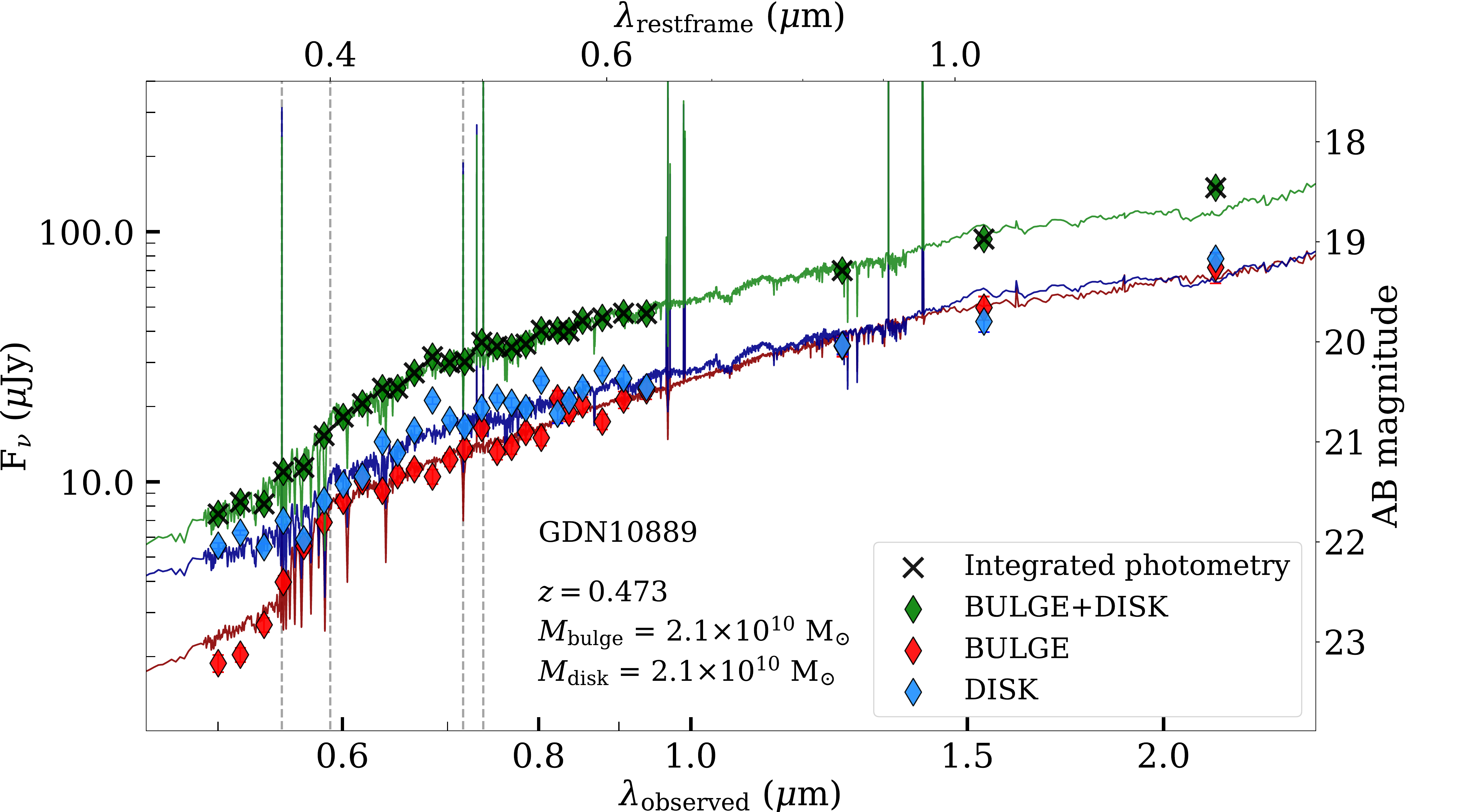}
\caption{Spectral energy distribution of the bulge (red), disk (blue), and galaxy (green) GDN~10889. 
Diamonds represent the individual photometric results of our decoupling analysis, while black crosses represent 
the measured integrated photometry of the galaxy in \citet{Barro2019}.
Errors are reported as 16th-84th percentile interval.
The best model for the bulge, disk, and galaxy are shown as red, blue, and green lines.
From left to right, vertical gray dashed lines represent the location of [OII], $D4000$, H$\beta$, and [OIII] features, respectively.
\label{fig:figure_2}}
\end{figure*}

In this second paper of a series we characterize the physical properties of the disk component of 
massive ($M_{\star} > 10^{10}$~M$_{\odot}$) and luminous ($m_{\rm F160W} <  21.5$ mag) 
galaxies at redshift $0.14 < z \leq 1$ in the North field of the Great Observatory Origins Deep Survey (GOODS-N).
Starting from the parent sample of 478 galaxies defined in \citetalias{Costantin2021a},
we consider the 91 galaxies with a reliable photometric bulge+disk decomposition (disky galaxies)
and lacking of any sign of interactions and/or background/foreground contaminating objects.
As discussed in \citetalias{Costantin2021a}, the representativeness of the sample in the SHARDS field-of-view is assured
both in redshift and stellar mass by means of a Kolmogorov-Smirnov test
(\mbox{$p$-value$_z > 10\%$}, \mbox{$p$-value$_{M_{\star}} > 30\%$};
see \citetalias{Costantin2021a}, for all details about the sample selection).

Our galaxies span a variety of bulge-over-total mass ratios ($0 \lesssim (B/T)_{\rm mass} \lesssim 1$),
bulge mass ($5 \times 10^{8} \lesssim M_{\rm b} \lesssim 2 \times 10^{11}$~M$_{\odot}$),
disk mass ($4\times10^{8} \lesssim M_{\rm d} \lesssim 10^{11}$~M$_{\odot}$),
bulge effective radii ($0.3 \lesssim R_{\rm e, b} \lesssim 5.6$~kpc), 
and disk effective radii ($2.4 \lesssim R_{\rm e, d} \lesssim 19$~kpc).
We present in Fig.~\ref{fig:figure_1} the distribution of $(B/T)_{\rm mass}$ and
stellar mass of the bulge and disk components.
Moreover, in Fig.~\ref{fig:figure_1} we visually demonstrate the large diversity of galaxy morphology
as a function of their $(B/T)_{\rm mass}$ and galaxy stellar mass.

\subsection{Spectrophotometric Decomposition \label{sec:section_2_2}}

For each galaxy and each filter, the bulge and disk light
is parametrized to be the sum of a \citet{Sersic1968}
and a single exponential function \citep{Freeman1970}, respectively.
We use the \texttt{GASP2D} algorithm \citep{MendezAbreu2008, MendezAbreu2014}
to obtain a two-dimensional model of the galaxy across wavelength.
Indeed, we took advantage of the 25 medium-band SHARDS images to
spectrophotometrically decouple the different stellar structures with a spectral resolution
$R\sim50$ in the wavelength range $0.500-0.941$~$\mu$m. 
We complement this information with the HST WFC3 and $K$-band data, 
covering the wavelength range between $\sim0.5$ and $2$~$\mu$m. 
It is worth remembering that the HST photometry is used as a prior 
for the decoupling of the bulge and disk light in SHARDS images
(see \citetalias{Costantin2021a}, for all details).
This strategy mimics the one used by the \texttt{C2D} code 
\citep{MendezAbreu2019a, MendezAbreu2019b},
which allows us to transfer the high spatial resolution
of the HST images to the SHARDS data set, consistently reducing the
degeneracies associated to the photometric modeling.
The procedure eventually provides us with the individual SEDs of each bulge and disk component.

\begin{deluxetable*}{cccccccccccc}
\tabletypesize{\scriptsize}
\tablecaption{Best parameters for the sample galaxies, bulges, and disks. \label{tab:table_1}}
\tablehead{
\colhead{ID}  & 
\colhead{$(B/T)_{\rm mass}$} &
\colhead{$R_{\rm e, b}$} & 
\colhead{$\log(M_{\rm b}$)} & 
\colhead{$\bar{t}_{M, \rm b}$} & 
\colhead{$\bar{z}_{M, \rm b}$} &
\colhead{log($\Sigma_{1.5, \rm b}$)} &
\colhead{$R_{\rm e, d}$} & 
\colhead{$\log(M_{\rm d}$)} & 
\colhead{$\bar{t}_{M, \rm d}$} & 
\colhead{$\bar{z}_{M, \rm d}$} &
\colhead{log($\Sigma_{1.5, \rm d}$)} \\
\colhead{} & 
\colhead{} & 
\colhead{(kpc)} & 
\colhead{(M$_{\odot}$)} & 
\colhead{(Gyr)} & 
\colhead{} &
\colhead{(M$_{\odot}$ kpc$^{-1.5}$)} &
\colhead{(kpc)} & 
\colhead{(M$_{\odot}$)} & 
\colhead{(Gyr)} & 
\colhead{} &
\colhead{(M$_{\odot}$ kpc$^{-1.5}$)}}
\decimalcolnumbers
\startdata
$750$  & $0.85 \pm 0.01$   &  $1.0 \pm 0.1$     & $10.64_{-0.07}^{+0.07}$ & $2.5_{-0.4}^{+0.3}$ & $1.8_{-0.2}^{+0.2}$ & $10.65 \pm 0.07$ & $6.7 \pm 0.4$ & $9.89_{-0.05}^{+0.05}$   & $1.2_{-0.2}^{+0.2}$ 	     & $1.27_{-0.05}^{+0.06}$ & $8.65 \pm 0.03$  \\
$775$  & $0.55 \pm 0.01$   &  $0.67 \pm 0.08$ & $10.72_{-0.09}^{+0.08}$ & $1.4_{-0.2}^{+0.1}$ & $1.4_{-0.1}^{+0.1}$ & $10.98 \pm 0.08$ & $5.7 \pm 0.4$ & $10.64_{-0.05}^{+0.05}$ & $3.6_{-0.5}^{+0.3}$       & $2.9_{-0.5}^{+0.5}$       & $9.51 \pm 0.05$  \\
$912$  & $0.72 \pm 0.01$   &  $0.79 \pm 0.04$ & $11.04_{-0.05}^{+0.06}$ & $3.7_{-0.6}^{+0.4}$ & $1.5_{-0.2}^{+0.2}$ & $11.20 \pm 0.03$ & $5.3 \pm 0.1$ & $10.64_{-0.05}^{+0.05}$ & $0.9_{-0.1}^{+0.1}$ 	     & $0.75_{-0.02}^{+0.03}$ & $9.56 \pm 0.02$  \\
$2104$ & $0.17 \pm 0.01$  &  $0.36 \pm 0.08$ & $10.09_{-0.09}^{+0.11}$ & $1.1_{-0.1}^{+0.2}$ & $1.0_{-0.1}^{+0.1}$ & $10.7 \pm 0.1$     & $3.6 \pm 0.1$ & $10.78_{-0.07}^{+0.07}$ & $3.6_{-0.7}^{+0.6}$ 	     & $2.1_{-0.4}^{+0.5}$       & $9.94 \pm 0.02$  \\
$5131$ & $0.92 \pm 0.01$  &  $0.9 \pm 0.2$     & $10.60_{-0.09}^{+0.07}$ & $5.7_{-0.7}^{+0.3}$ & $7.3_{-3.1}^{+4.2}$ & $10.7 \pm 0.1$     & $2.6 \pm 0.2$ & $9.53_{-0.05}^{+0.05}$   & $0.16_{-0.02}^{+0.03}$ & $0.88_{-0.01}^{+0.01}$ & $8.91 \pm 0.04$  \\
\enddata
\tablecomments{Table \ref{tab:table_1} is published in its entirety in the machine-readable format.
 A portion is shown here for guidance regarding its form and content.
Column (1): CANDELS ID of the galaxy \citep{Barro2019}. Column (2): mass-weighted $B/T$.
Column (3): bulge effective radius at 1.6~$\mu$m (\citetalias{Costantin2021a}). Column (4): bulge stellar mass.
Column (5): bulge mass-weighted age. Column (6): bulge mass-weighted formation redshift.
Column (7): bulge mass surface density.
Column (8): disk effective radius at 1.6~$\mu$m (\citetalias{Costantin2021a}). Column (9): disk stellar mass.
Column (10): disk mass-weighted age. Column (11): disk mass-weighted formation redshift.
Column (12): disk mass surface density.}
\end{deluxetable*}

\subsection{Stellar Populations \label{sec:section_2_3}}

We fit both the measured bulge and disk SEDs
with the \citet{Bruzual2003} stellar population library
by means of the \texttt{synthesizer} fitting code 
\citep[see][for all details]{PerezGonzalez2003, PerezGonzalez2008}.
Shortly, we assume a \citet{Chabrier2003} initial mass function 
integrated in the range $0.1 < M/M_{\odot} < 100$.
The star formation history (SFH) of each galaxy component is parametrized 
with a declining delayed exponential law
\begin{equation}
SFR(t) \propto t/\tau^2 \, e^{-t/\tau} \, ,
\end{equation}
where $\tau$ runs from 200 Myr
to a roughly constant SFH ($\tau = 100$ Gyr). 
The metallicity of the models spans discrete values $Z/Z_{\odot}$ = [0.4, 1, 2.5] 
(i.e., sub-solar, solar, and super-solar). 
The extinction law of \citet{Calzetti2000} is used to 
parametrize the V-band attenuation, with values ranging from 0 to 3 mag.

As described in \citetalias{Costantin2021a}, the $\chi^2$ maximum-likelihood estimator is minimized to
obtain the best fitting model. Moreover, for each galaxy component
we run 500 Monte Carlo simulations to estimate
the uncertainties in the stellar population parameters
and to account for possible degeneracies in the solutions
\cite[see][for more details]{DominguezSanchez2016}.
As an example, in Fig.~\ref{fig:figure_2} we present the best model for the bulge, 
disk, and galaxy SED for the galaxy GDN~10889. 

\begin{figure*}[t!]
\centering
\includegraphics[scale=0.4, trim=0cm 0cm 0cm 0cm , clip=true]{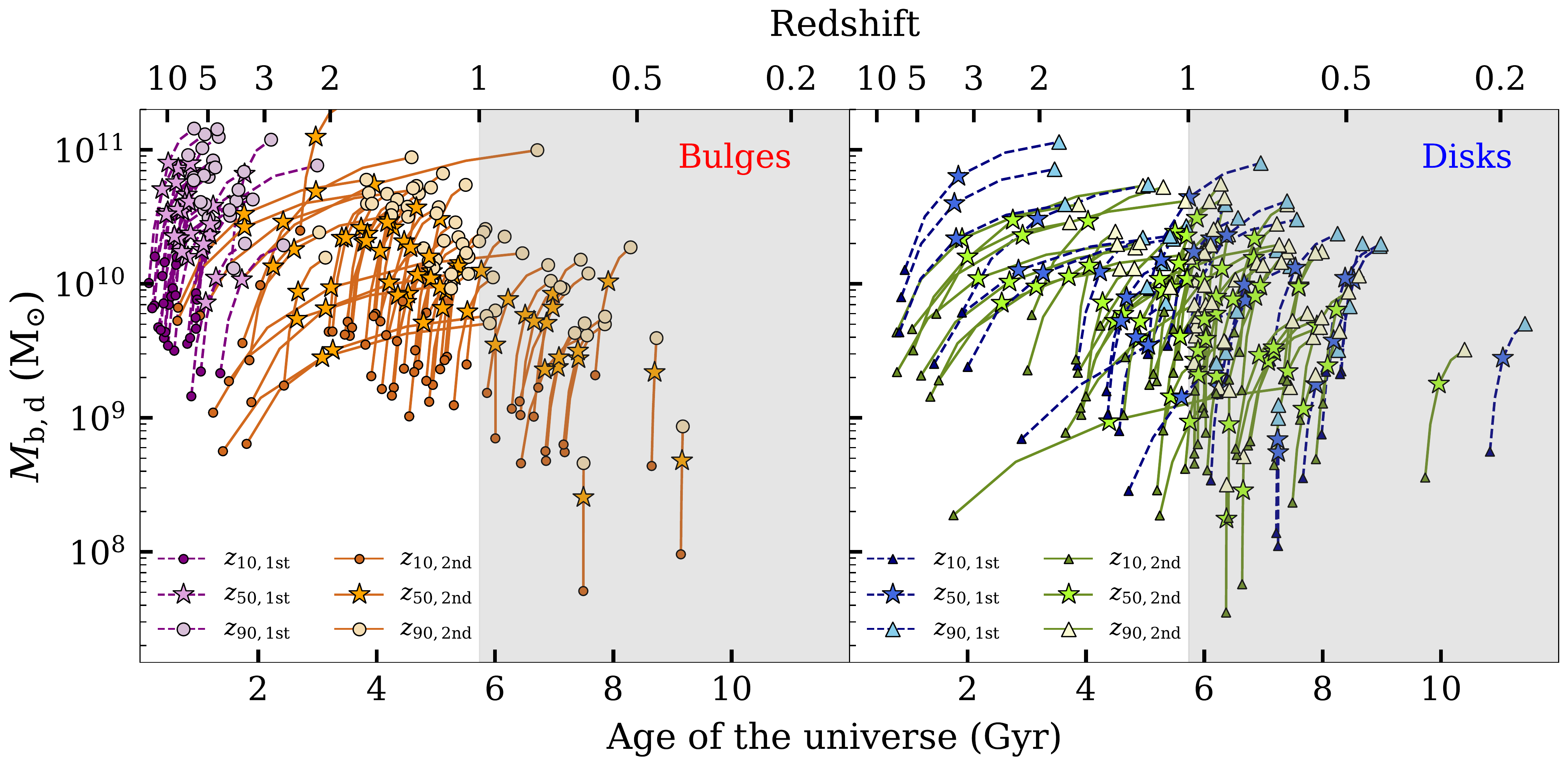}
\caption{Mass assembly history of our bulges (left panel) and disks (right panel)
as a function of the age of the universe.
Bulges are separated between first-wave (purple dashed lines) and second-wave ones (orange solid lines).
Disks are separated between those around first-wave bulges (blue dashed lines) 
and those around second-wave bulges (green solid lines).
For each system we mark the instants when 
they build 10\%, 50\%, and 90\% of the current stellar mass (from darker to lighter colors, from smaller to larger sizes).
In particular, stars stand for $z_{50}$, i.e., the redshift when each system grows half of its current mass.
The gray shaded regions stand for the redshift of observation.
\label{fig:figure_3}}
\end{figure*}

The characterization of each bulge and disk SFH
allows us to derive fundamental physical quantities, which constrain their stellar populations
and result critical to quantify their evolutionary process.
In particular, in this paper we will characterize the bulges and disks of the galaxies in our sample 
in terms of the following physical properties:
the stellar mass ($M_{\star}$), the star formation timescale ($\tau$), 
the metallicity ($Z$), and the dust attenuation ($A_{\rm V}$).
Furthermore, we compute the bulge and disk mass-weighted age ($\bar{t}_{M}$)
as well as its corresponding redshift (i.e., the mass-weighted formation redshift $\bar{z}_{M}$).
While mass-weighted ages take into account the extent of the star formation
and mitigate the age-$\tau$ degeneracy,
the mass-weighted formation redshift allows us to properly compare galaxies observed at different redshift.
In order to characterize the beginning and the end of the mass assembly of our bulges and disks,
we calculate the cosmic times corresponding to the instants when they acquire a fraction of their current mass,
starting from the onset of the first episode of star formation.
In particular, considering their observed redshift, we convert each instant to the
corresponding redshift $z_{10}$, $z_{50}$, and $z_{90}$,
i.e., the redshift when each component grows 
10, 50, and 90 per cent of its current mass 
(see Fig.~\ref{fig:figure_3}).
With this definition $z_{50}$ is a proxy for the mass-weighted formation redshift $\bar{z}_{M}$
and they can be compared at first approximation (see Sect.~\ref{sec:section_3_3}).
Finally, we also derive the bulge and disk compactness, 
computing their mass surface density $\Sigma_{1.5} = M R_{\rm e}^{-1.5}$ \citep{Barro2013}.
We report in Table~\ref{tab:table_1} the properties derived from the 
main cluster of solutions for the sample galaxies, bulges, 
and disks, respectively.


\section{Results \label{sec:section_3}}

In this paper we focus on the characterization 
of the stellar disks in a representative sample of massive galaxies at redshift $0.14 < z \leq 1$,
comparing their properties with those of bulges presented in \citetalias{Costantin2021a}.
In particular, we present the bulge and disk mass assembly
and star formation histories in Sect.~\ref{sec:section_3_1} and \ref{sec:section_3_2},
their individual mass-weighted formation redshifts in Sect.~\ref{sec:section_3_3},
and the interplay between their ages and fundamental physical properties
(i.e., mass, size, star formation timescale, mass surface density, 
and S\'ersic index) in Sect.~\ref{sec:section_3_4}.

\subsection{Bulges and Disks Mass Assembly History \label{sec:section_3_1}}

In this Section we describe how bulges and disks build their mass,
focusing on the timescales of their SFHs. This will allow us to understand 
which component starts to build earlier and at which rate they form. 

In Fig.~\ref{fig:figure_3} we show the stellar masses 
of our bulges and disks as a function of the age of the universe,
i.e., the evolutionary tracks in their stellar mass assembly.
We identify (left panel) two behaviors
when considering first-wave bulges (those with mass-weighted formation redshift beyond $z=3$) and
second-wave bulges (those formed from $z=3$), as identified in \citetalias{Costantin2021a}.
Indeed, first-wave bulges grow their mass on short timescales $t_{90} - t_{10} = 0.68^{+0.05}_{-0.01}$~Gyr;
almost identical values are found for those second-wave bulges which start to assemble at redshift
$z \lesssim 2$ ($t_{90} - t_{10} = 0.68^{+0.03}_{-0.33}$~Gyr). 
On the other hand, there is a fraction of
second-wave bulges (14\%) that start to assemble at $2 \lesssim z \lesssim 5$,
build half of their mass at redshift $z  < 3$, and evolve more
slowly $t_{90} - t_{10} = 3.4^{+1.2}_{-1.0}$~Gyr.
This subpopulation corresponds to more horizontal evolutionary tracks in the panel.

In Fig.~\ref{fig:figure_3} (right panel) we see that some disks start to form as early as 
first-wave bulges ($z \sim 5$). However, they assemble more
slowly than bulges ($t_{90} - t_{10} = 3.5^{+0.9}_{-0.8}$~Gyr), accreting half of their current mass
by redshift $z \sim 2-4$. 
This can be understood if a fraction of the primordial gaseous disk forms clumps that migrate to the bulge.
In this scenario, only a fraction of the new stars formed in the disk remains in the disk.
On the other hand, the evolution of disks which start to form 
at redshift $z \lesssim 1.5$ is faster ($t_{90} - t_{10} = 0.7^{+0.5}_{-0.4}$~Gyr). 
However, it is worth noticing that 21 out of 66 (32\%) of those disks present significant 
ongoing star formation and/or assembly activity ($\bar{t}_{M} < 500$~Myr) at the redshift of observation (see Fig.~\ref{fig:figure_A}). 
Thus, given that these disks are young, the star formation timescale is not very informative; in principle, they 
could eventually present similar timescales to those of the disks formed at higher redshift 
(which we catch in a more evolved evolutionary stage).

We find that the bulk of our disks (76\%) grows half of their mass
later than bulges despite some disks having started to form as early as some bulges ($z\sim5$).
Therefore, even if bulges and disks could start forming at the same time, 
bulges assemble rapidly (short $\tau_{\rm b}$) and then evolve passively, 
but disks continue forming stars, so they continue growing more slowly.
We note that the first-wave bulges present very different timescales
compared to the disks around them. The formation of these structures are far enough 
from the time corresponding to the observed redshift that we can analyze 
the long-term evolution of bulges and disks. Interestingly, 
bulges quench quite rapidly, while disks present more extended star formation (see Sect.~\ref{sec:section_3_2}). 
This indicates that a large amount of gas still exists and falls into the galaxy, 
but it is not transferred to the central region and/or cannot be transformed efficiently into stars.

\begin{figure}[t!]
\centering
\includegraphics[scale=0.36]{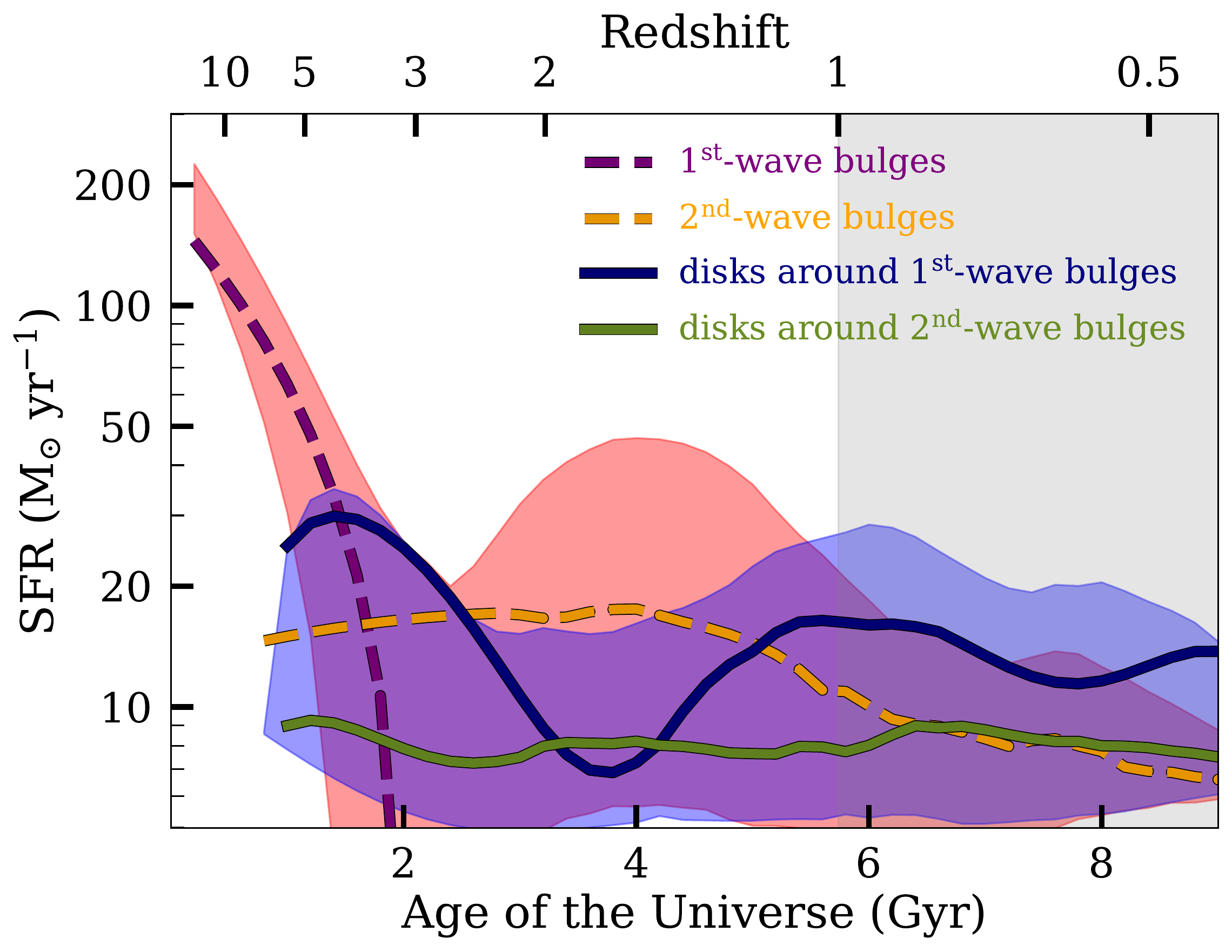}
\caption{Averaged SFHs of the first-wave bulges (purple dashed line)
and disks around them (blue solid line), compared with the one of
second-wave bulges (orange dashed line) and disks around them (green solid line).
The red and blue shaded curves represent the 16th--84th percentile interval
computed from the scatter of the SFHs of bulges and disks, respectively.
The gray shaded area indicates the redshift studied in this work.
\label{fig:figure_4}}
\end{figure}

\begin{figure*}[t!]
\centering
\includegraphics[scale=0.5, trim=0cm 0cm 0cm 0cm , clip=true]{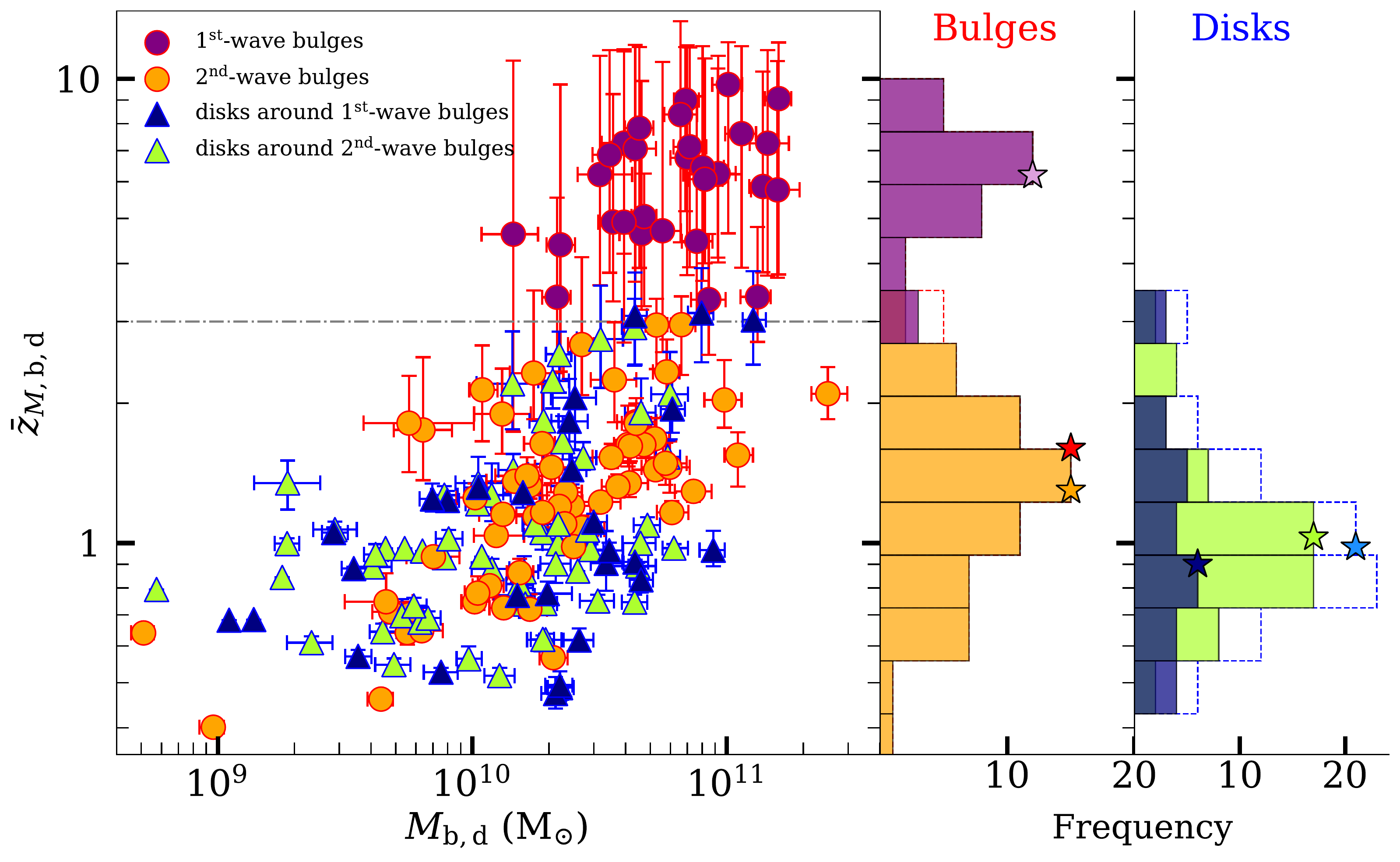}
\caption{Mass-weighted formation redshift of bulges (dots) and disks (triangles) as a function
of their stellar mass. Bulges and disks are separated in 
first-wave (purple and blue, respectively) and 
second-wave ones (orange and green, respectively).
Errors are reported as 16th--84th percentile interval.
The gray dash-dotted horizontal line marks $\bar{z}_{M} = 3$.
The histograms represent the frequency of the mass-weighted formation redshifts
of the bulge and disk populations, respectively. 
Purple and orange histograms stand for first and second-wave bulges,
while the histogram with red dashed contour stands for the entire bulge population.
Blue and green histograms stand for disks around first and second-wave bulges,
while the histogram with light blue dashed contour stands for the entire disk population.
The median values of each distribution are marked with stars.
\label{fig:figure_5}}
\end{figure*}

\subsection{Bulges and Disks Star Formation History \label{sec:section_3_2}}

As discussed in \citetalias{Costantin2021a}, a considerable fraction of our massive galaxies 
experienced a peak of star formation at $\bar{z}_{M, \rm b} > 5$ (see Fig.~\ref{fig:figure_4}), linked to a violent
episode of compaction (revealed by the high stellar mass densities).
The large SFR values which led to the formation of first-wave bulges
are consistent with the high gas accretion rates and the successive compaction events
\citep{Zolotov2015, Ceverino2018}. In particular, \citet{Ceverino2018} reported typical star formation bursts at $z=10$ with maximum 
specific SFR of $\sim 20$~Gyr$^{-1}$, which translates to $SFR \sim 200$~M$_{\odot}$~yr$^{-1}$ for $10^{10}$ M$_{\odot}$.
This episode causes the rapid grow of a small but massive system,
which evolves as a compact spheroid (bulge) and
develops an extended (and dynamically stable) stellar disk at later times, as we show in this paper.
Thus, we identify these bulges as relics of the blue and red-nugget population 
usually observed at $z\gtrsim1.5$ \citep{Damjanov2009, Barro2013}.

We showed in Fig.~\ref{fig:figure_3} that the first disks started to form at redshift $z \sim 5$.
But, they build up on longer timescales with respect to bulges, assembling
half of their current mass by redshift $z \sim 2$.
They are characterized by long timescales of
evolution ($\tau_{\rm d} > 1$~Gyr), small sizes ($R_{\rm e, d} \sim 4$~kpc), and high mass surface densities
(log$(\Sigma_{1.5, \rm d}) \gtrsim 9.5$~M$_{\odot}$~kpc$^{-1.5}$). 
The compactness of these disks and of first-wave bulges  
suggests that in the $z \sim 3$ universe the conditions 
were favorable for shaping small and massive systems,
but with different morphologies.

As shown in Fig.~\ref{fig:figure_4}, the episode of star formation for disks which develop around first-wave bulges
is on average more intense than the one of disks around second-wave bulges ($SFRs$ of $\sim 20-30$ and 
$\sim 10$~M$_{\odot}$~yr$^{-1}$, respectively).
As the cosmic time passes ($z \sim 3$ and lower), a second wave of spheroids 
starts to assemble. These second-wave bulges present a 
variety of timescales, but the majority of them still form on short 
timescales ($\tau_{\rm b} < 500$~Myr) with peaks of star formation
as intense as $\sim 50$~M$_{\odot}$~yr$^{-1}$ (and higher than coeval disks).
Nevertheless, on average bulges and disks form stars at similar rates
from redshift $z \sim 1.5$ to redshift $z \sim 1$.
At redshift $z\sim1$ our bulges display a drop of star formation,
while disks show a slight increase of star formation activity (until the time of observation).

Summarizing our results in this subsection, we identify four morphological epochs in the evolution of massive disk galaxies:
a first wave of bulges started to assemble as early as redshift $z\sim10$ ($\sim 13$~Gyr ago); 
after some time ($\sim12.3$~Gyr ago; $z\sim5$) the first population (15\%) of extended stellar disks start to assemble,
but on longer timescales than the first wave of bulges.
Then, a second wave of bulges, predominant in number ($67\%$),
starts to grow from redshift $z\sim2.5$ down to redshift $z\sim1$. On the top of this second wave, 
starting from redshift $z\sim1.5$ ($\sim 9.3$~Gyr ago), the disk era starts.
Even though bulges are on average older than disks (see Sect.~\ref{sec:section_3_3_2}),
we find a close interplay between the two components,
which suggest a level of co-evolution between them. This will be further discussed in Sect.~\ref{sec:section_4}.

\begin{figure*}[t!]
\centering
\includegraphics[scale=0.5, trim=0cm 0cm 0cm 0cm , clip=true]{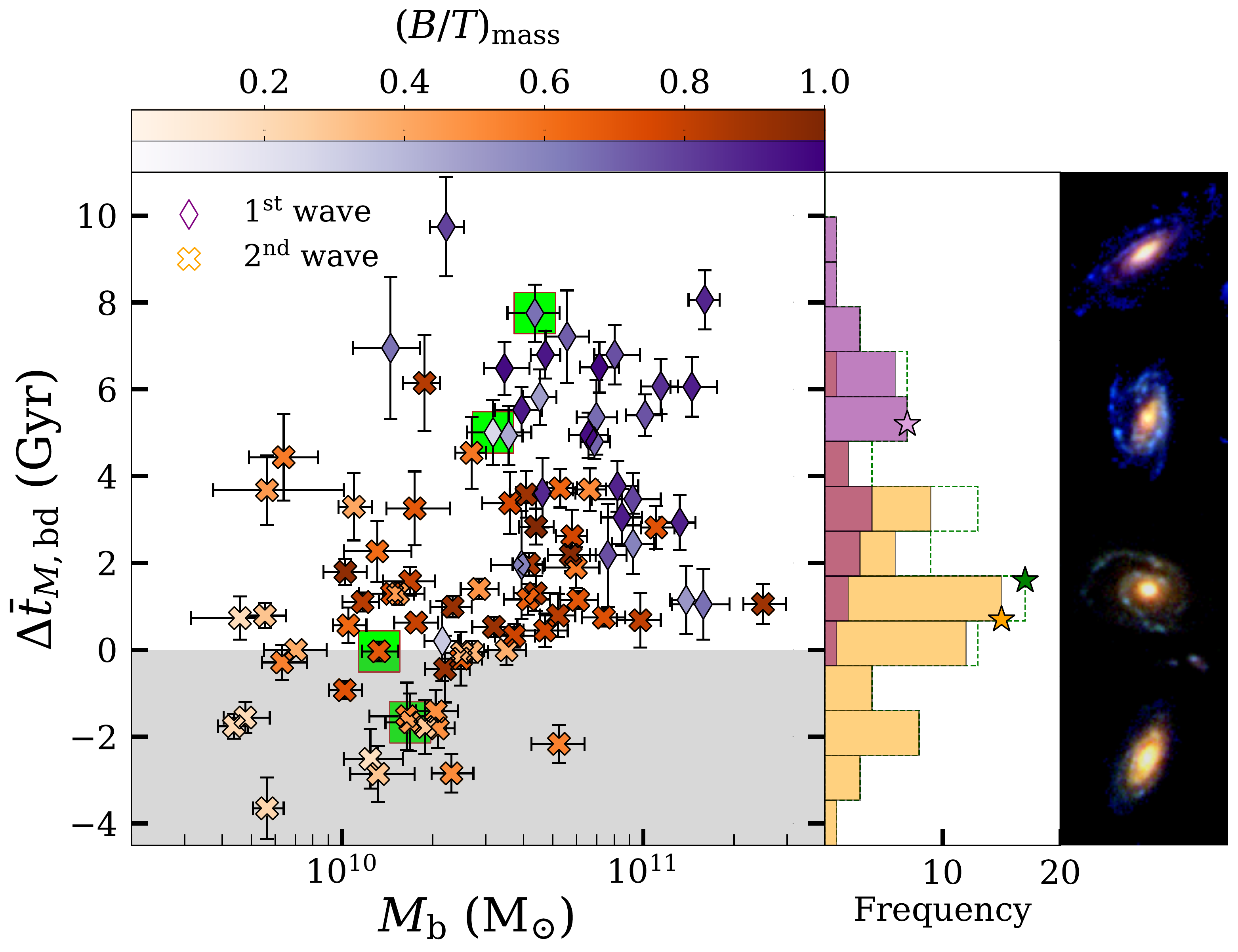}
\caption{Age difference between each bulge and disk as a function
of the bulge mass, color-coded according to $(B/T)_{\rm mass}$.
Errors are reported as the 16th--84th percentile interval.
The gray shaded region marks $\Delta \bar{t}_{M, \rm bd} < 0$.
Green squares mark the four galaxies shown in RGB colors as an example (right panels).
The purple histogram stands for galaxies with first-wave bulges,
the orange histogram stands for galaxies with second-wave bulges,
and the histogram with green dashed contour stands for the entire population.
The median values of each distribution are marked with stars.
\label{fig:figure_6}}
\end{figure*}

\subsection{Mass-weighted Formation Redshifts \label{sec:section_3_3}}

We derive the mass-weighted ages of each bulge and disk,
providing fundamental constraints on their formation and co-evolution 
(see Table~\ref{tab:table_1} and Appendix~\ref{app:A}).
Our galaxies are observed in a wide redshift range $0.14 < z \leq 1$ (spanning $\sim6$~Gyr of cosmic time),
making the interpretation of their mass-weighted ages more difficult.
Thus, we compute the redshift corresponding to
their mass-weighted ages ($\bar{z}_{M}$; see Table~\ref{tab:table_1})
to follow the evolution of each system at different epochs.
In Fig.~\ref{fig:figure_5} we show the mass-weighted formation redshift of bulges and disks
as a function of their stellar mass.
As already discussed in Sect.~\ref{sec:section_3_1}, 
we find that the disk population forms on average at later cosmic times than the bulge population.
Bulges have median mass-weighted formation redshift $\bar{z}_{M, \rm b} = 1.6^{+4.6}_{-0.7}$,
while disks have $\bar{z}_{M, \rm d} = 1.0^{+0.6}_{-0.3}$.
In \citetalias{Costantin2021a} we found that 33\% of bulges have $\bar{z}_{M, \rm b} > 3$.
Here we show that only ten out of 91 disks have $\bar{z}_{M, \rm d}> 2$ and
none of them presents $\bar{z}_{M, \rm d} > 3.2$.
These findings are consistent with the predictions from multiple
cosmological simulations, where the spheroidal component tends to form at early cosmic epochs
and late star formation contributes to the growth of disk stars.
In particular, \citet{Park2019} quantified that massive galaxies ($10 <$ log($M_{z=0.7}$/M$_{\odot}) < 11$) 
in the New Horizon simulations \citep{Dubois2021} start to form disks from
$z\sim1-2$, once their mass become $\sim10^{10}~$M$_{\odot}$.
\citet{Tacchella2019} showed that the efficiency of disk formation in Illustris TNG simulations \citep{Pillepich2018,Nelson2018}
strongly depends on both on stellar mass and cosmic time. 
In both simulations, at early times the formation efficiency of disks is low, increasing towards $z\sim1$.

Bulges formed at higher redshift are more massive
(\citetalias{Costantin2021a}), as proved by a Spearman correlation coefficient
of 0.71 ($p$-value $< 0.01$). Similarly, older disks are more massive
(Spearman correlation coefficient of 0.40; $p$-value $< 0.01$).
On the other hand, the bulge formation time has a stronger correlation with the total mass
of the galaxy than disks, having a Spearman correlation 
coefficient of 0.63 and 0.36, respectively.
These trends are in agreement with the fact that there is a strong correlation between 
mass and cosmic time, as shown both in the FIRE \citep{Hopkins2014} 
and the New Horizon cosmological simulations.
In particular, massive galaxies start to form disk stars at earlier epochs ($z\gtrsim1$),
while low-mass galaxies develop their disks after $z\sim1$ \citep{ElBadry2018, Park2019}.

\subsubsection{First and Second-wave Systems \label{sec:section_3_3_1}}

As discussed in \citetalias{Costantin2021a}, bulges form in two waves.
The distribution of their mass-weighted formation redshift is bimodal: first-wave bulges 
have median mass-weighted formation redshift $\bar{z}_{M, \rm b} = 6.2^{+1.5}_{-1.7}$,
while second-wave bulges have median mass-weighted formation redshift \mbox{$\bar{z}_{M, \rm b} = 1.3^{+0.6}_{-0.6}$}.
Our analysis allows us to directly compare the disks which build around first-wave bulges
with the ones growing around second-wave bulges.
The first ones have median mass-weighted formation redshift $\bar{z}_{M, \rm d} = 0.9^{+0.9}_{-0.4}$, 
while the latter ones have $\bar{z}_{M, \rm d} = 1.0^{+0.6}_{-0.3}$.
A Kolmogorov-Smirnov test ($K = 0.15$, $p$-value $> 0.6$) suggests that 
we cannot reject the null hypothesis that the two $\bar{z}_{M, \rm d}$ distributions are identical.
Thus, at the peak of the cosmic SFR density \citep{Lilly1996, Madau1996, Madau2014}
disks could form around all types of spheroids. 

We find that most of our disks formed at similar cosmic times than second-wave bulges 
($\bar{z}_{M} \sim 1$).
Nonetheless, the disks $\bar{z}_{M, \rm d}$ (and $\bar{t}_{M,\rm d}$) distribution
is different from the one of second-wave bulges
at $3\sigma$ confidence level, as proved by a Kolmogorov-Smirnov 
test with statistic $K = 0.35$ ($K = 0.30$).

\subsubsection{Age Difference between Bulges and Disks \label{sec:section_3_3_2}}

In order to characterize the galaxy evolution in terms of its structural components,
we show in Fig.~\ref{fig:figure_6} the differences in mass-weighted ages between each bulge and disk.
The median age difference is $\Delta\bar{t}_{M, \rm bd} = \bar{t}_{M, \rm b} - \bar{t}_{M, \rm d}  = 1.6^{+5.4}_{-0.7}$~Gyr.

We define bulges as older, coeval, and younger than their disks by looking
at the compatibility between each 
$\Delta\bar{t}_{M, \rm bd} \pm \sigma_{\Delta \bar{t}_{M, \rm bd}}$ and $\Delta\bar{t}_{M, \rm bd}=0$.
At $1\sigma$ ($3\sigma$) level we find that 74\% (59\%) of our bulges formed before their disks,
9\% (29\%) of systems are compatible to be coeval,
while in 17\% (12\%) of galaxies the bulge formed after the disk component.
In particular, all first-wave bulges are older than their disk component.
We find that first-wave bulges and their disks present 
$\Delta \bar{t}_{M, \rm bd} = 5.2^{+1.1}_{-1.9}$~Gyr, while the age difference 
for second-wave systems is  $\Delta \bar{t}_{M, \rm bd} = 0.7^{+1.5}_{-1.6}$~Gyr.
In the following, we analyze the properties of bulges and disks 
considering the age difference between the two components at $1\sigma$ level.

As discussed in Sect.~\ref{sec:section_4}, 
our results actually suggest that (second-wave) bulges and disks
display a degree of co-evolution: both structures form at similar times ($z_{10}$), but  
the disk builds up on longer timescales, keeping on forming stars while the bulge stops
its star formation earlier (see Figs.~\ref{fig:figure_3} and \ref{fig:figure_4}).

\subsection{Morphological Properties \label{sec:section_3_4}}

In this Section we analyze which are the physical properties
regulating the bulge and disk evolution. In particular,
we focus on their stellar mass, size, timescale, mass surface density, 
and S\'ersic index.

\subsubsection{Stellar Mass \label{sec:section_3_4_1}}

Our bulges have median 
$\log (M_{\rm b}/{\rm M}_{\odot}) = 10.5^{+0.3}_{-0.5}$,
while our disks have median $\log (M_{\rm d}/{\rm M}_{\odot}) = 10.2^{+0.4}_{-0.6}$.

We find that first and second-wave bulges have median 
$\log (M_{\rm b}/{\rm M}_{\odot}) = 10.8^{+0.2}_{-0.3}$ and
\mbox{$\log (M_{\rm b}/{\rm M}_{\odot}) = 10.3^{+0.4}_{-0.5}$}, respectively.
On the other hand, disks around first and second-wave bulges have similar masses of 
$\log (M_{\rm d}/{\rm M}_{\odot}) = 10.3^{+0.3}_{-0.7}$ and
$\log (M_{\rm d}/{\rm M}_{\odot}) = 10.2^{+0.3}_{-0.5}$, respectively.
A Kolmogorov-Smirnov test suggests 
that we cannot reject the null hypothesis 
that both the two $M_{\rm d}$ distributions ($K = 0.27$, $p$-value $> 0.08$) 
and the distributions of $M_{\rm d}$ and $M_{\rm b}$ for
second-wave bulges ($K = 0.17$, $p$-value $> 0.2$)  are similar.

\begin{figure}[t!]
\centering
\includegraphics[scale=0.34]{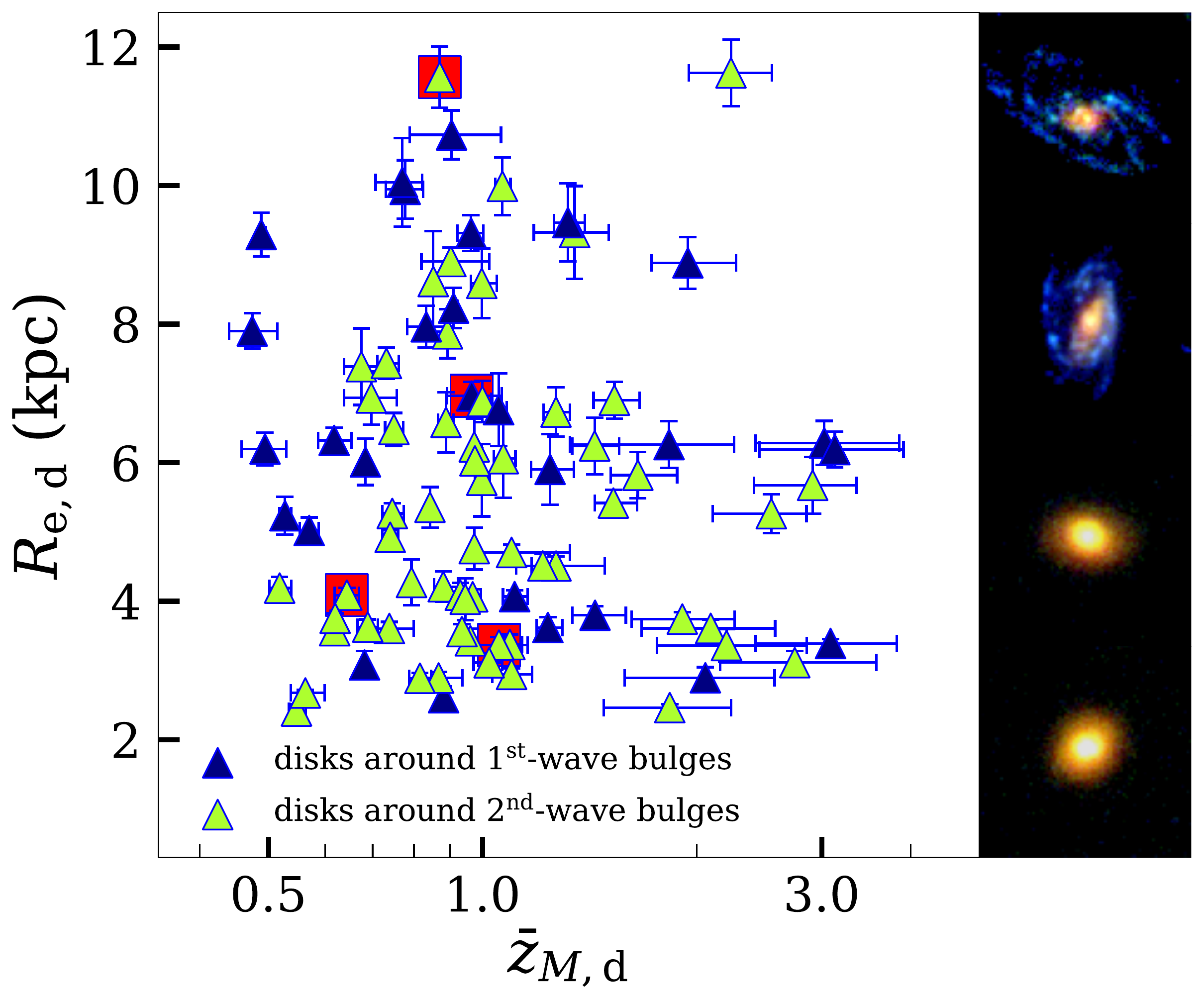}
\caption{Size of disks as a function of their mass-weighted formation redshift.
Disks around first and second-wave bulges
are shown in blue and green, respectively.
Errors are reported as the 16th--84th percentile interval.
Red squares mark the four galaxies shown in
RGB colors as an example (right panels).
\label{fig:figure_7}}
\end{figure}

In terms of their mass, bulges older, coeval, and younger than their disks have median mass 
$\log (M_{\rm b}/{\rm M}_{\odot}) = 10.6^{+0.3}_{-0.5}$,
$\log (M_{\rm b}/{\rm M}_{\odot}) = 10.4^{+0.1}_{-0.2}$,
and $\log (M_{\rm b}/{\rm M}_{\odot}) = 10.1^{+0.2}_{-0.5}$, respectively.
There is a (mild) correlation (Spearman coefficient of 0.48; $p$-value $< 0.01$) between
$\Delta \bar{t}_{M, \rm bd}$ and $M_{\rm b}$.
In particular, in the high mass regime ($M_{\rm b} > 3 \times 10^{10}$~M$_{\odot}$)
there is only one bulge (out of 54) younger than its disk (see Fig.~\ref{fig:figure_6}).
Disks younger, coeval, and older than their bulges have median mass 
$\log (M_{\rm d}/{\rm M}_{\odot}) = 10.2^{+0.3}_{-0.6}$,
$\log (M_{\rm d}/{\rm M}_{\odot}) = 10.4^{+0.3}_{-0.6}$,
and $\log (M_{\rm d}/{\rm M}_{\odot}) = 10.3^{+0.3}_{-0.2}$, respectively.
There is no correlation between $\Delta \bar{t}_{M, \rm bd}$ and $M_{\rm d}$.

We investigate now the relative contribution of bulges and disks to the total stellar mass of our galaxies.
The median bulge-over-total mass ratio of the sample is $(B/T)_{\rm mass} = 0.69^{+0.21}_{-0.31}$.
Galaxies which build from first-wave systems are more bulge-dominated
than those which build later on, having $(B/T)_{\rm mass} = 0.76^{+0.16}_{-0.18}$ and
$(B/T)_{\rm mass} = 0.63^{+0.22}_{-0.29}$, respectively.
There is a mild correlation between the bulge prominence and its formation redshift
(Spearman coefficient of 0.45; $p$-value $< 0.01$), while we find no correlation
between the disk $\bar{z}_{M, \rm d}$ and $(B/T)_{\rm mass}$.

We find a trend between $\Delta \bar{t}_{M, \rm bd}$ and the prominence of the two components
defined by $(B/T)_{\rm mass}$,
as shown in Fig.~\ref{fig:figure_6}. 
The Spearman correlation coefficient of ($\Delta \bar{t}_{M, \rm bd}$, $(B/T)_{\rm mass}$) 
is 0.49 ($p$-value $< 0.01$).
We find that 48\% of bulges in galaxies with $B/T<0.5$ are younger than their disks 
(12 out of 25), while this fraction diminishes to 6\% for bulges in galaxies with $B/T>0.5$.
In particular, bulges older than their disks reside in galaxies with median 
$(B/T)_{\rm mass} = 0.75^{+0.16}_{-0.20}$, bulges with ages similar of those 
of their disks reside in galaxies with median $(B/T)_{\rm mass} = 0.46^{+0.23}_{-0.11}$, 
and bulges younger than their disks are found in galaxies with median 
$(B/T)_{\rm mass} = 0.34^{+0.18}_{-0.16}$.

\begin{figure}[t!]
\centering
\includegraphics[scale=0.34]{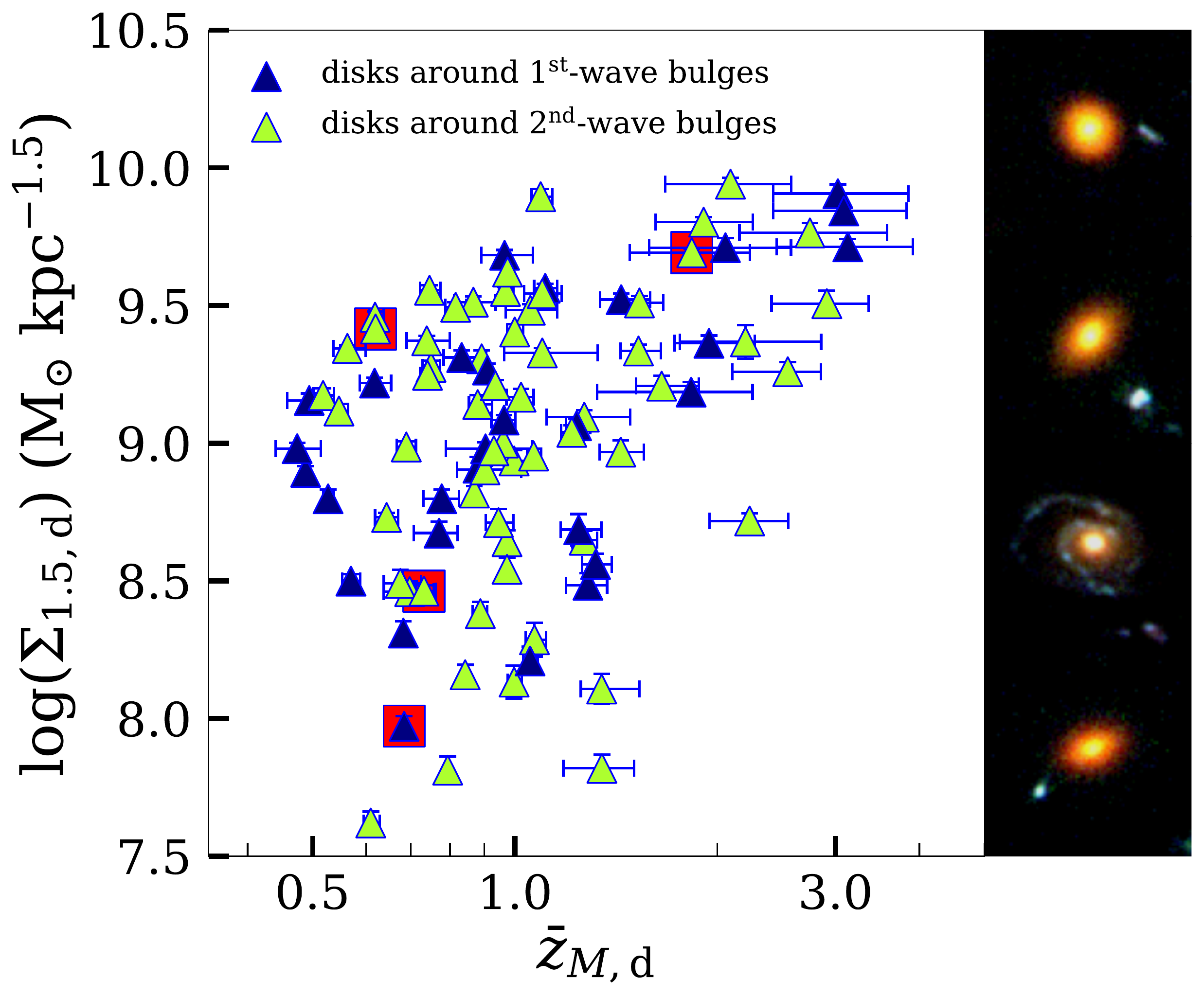}
\caption{Mass surface density of disks as a function of their mass-weighted formation redshift.
Disks around first and second-wave bulges
are shown in blue and green, respectively.
Errors are reported as 16th--84th percentile interval.
Red squares mark the four galaxies shown in
RGB colors as an example (right panels).
\label{fig:figure_8}}
\end{figure}

\subsubsection{Size \label{sec:section_3_4_2}}

We define the size by means of the effective radius $R_{\rm e}$,
i.e., the radius at which bulges and disks contain half of their light in the WFC3 F160W band.
In particular, being the surface brightness of the disk modeled with 
an exponential profile with scale radius $h$ (\citetalias{Costantin2021a}),
the effective radius corresponds to $R_{\rm e, d} = 1.678 \times h$ \citep{Graham2005}.

Bulges and disks have median sizes $R_{\rm e, b} = 1.0^{+0.9}_{-0.4}$~kpc
and $R_{\rm e, d} = 5.4^{+3.4}_{-2.0}$~kpc, respectively.
We find that first and second-wave bulges have similar median sizes, i.e.,
$R_{\rm e, b} = 1.3^{+0.8}_{-0.6}$~kpc and $R_{\rm e, b} = 1.0^{+0.8}_{-0.4}$~kpc, respectively.
On the other hand, we show in Fig.~\ref{fig:figure_7} that 
disks around first and second-wave bulges have
$R_{\rm e, d} = 6.3^{+3.1}_{-2.5}$~kpc and
$R_{\rm e, d} = 4.8^{+2.8}_{-1.4}$~kpc, respectively.
Disks around first-wave bulges are larger than those around second-wave bulges
at $2\sigma$ confidence level ($K = 0.31$).
In \citetalias{Costantin2021a} we found a weak correlation (Spearman coefficient of 0.21) between the mass-weighted
formation redshift of bulges and their size. However, there is no correlation between the disk size 
and the time of their formation (Fig.~\ref{fig:figure_7}). This remains valid if we separate disks in those 
around first and second-wave bulges.

In terms of their size bulges older, coeval, and younger than their disks have median sizes 
$R_{\rm e, b} = 1.1^{+0.9}_{-0.4}$~kpc,
$R_{\rm e, b} = 1.0^{+0.4}_{-0.2}$~kpc,
and $R_{\rm e, b} = 0.7^{+0.7}_{-0.3}$~kpc, respectively.
Disks younger, coeval, and older than their bulges have median sizes 
$R_{\rm e, d} = 6.0^{+3.1}_{-2.4}$~kpc,
$R_{\rm e, d} = 6.9^{+0.5}_{-3.4}$~kpc,
and $R_{\rm e, d} = 3.6^{+1.9}_{-0.6}$~kpc, respectively.
There is no correlation between neither $\Delta \bar{t}_{M, \rm bd}$ and $R_{\rm e, b}$
nor $\Delta \bar{t}_{M, \rm bd}$ and $R_{\rm e, d}$.

\begin{figure}[t!]
\centering
\includegraphics[scale=0.31]{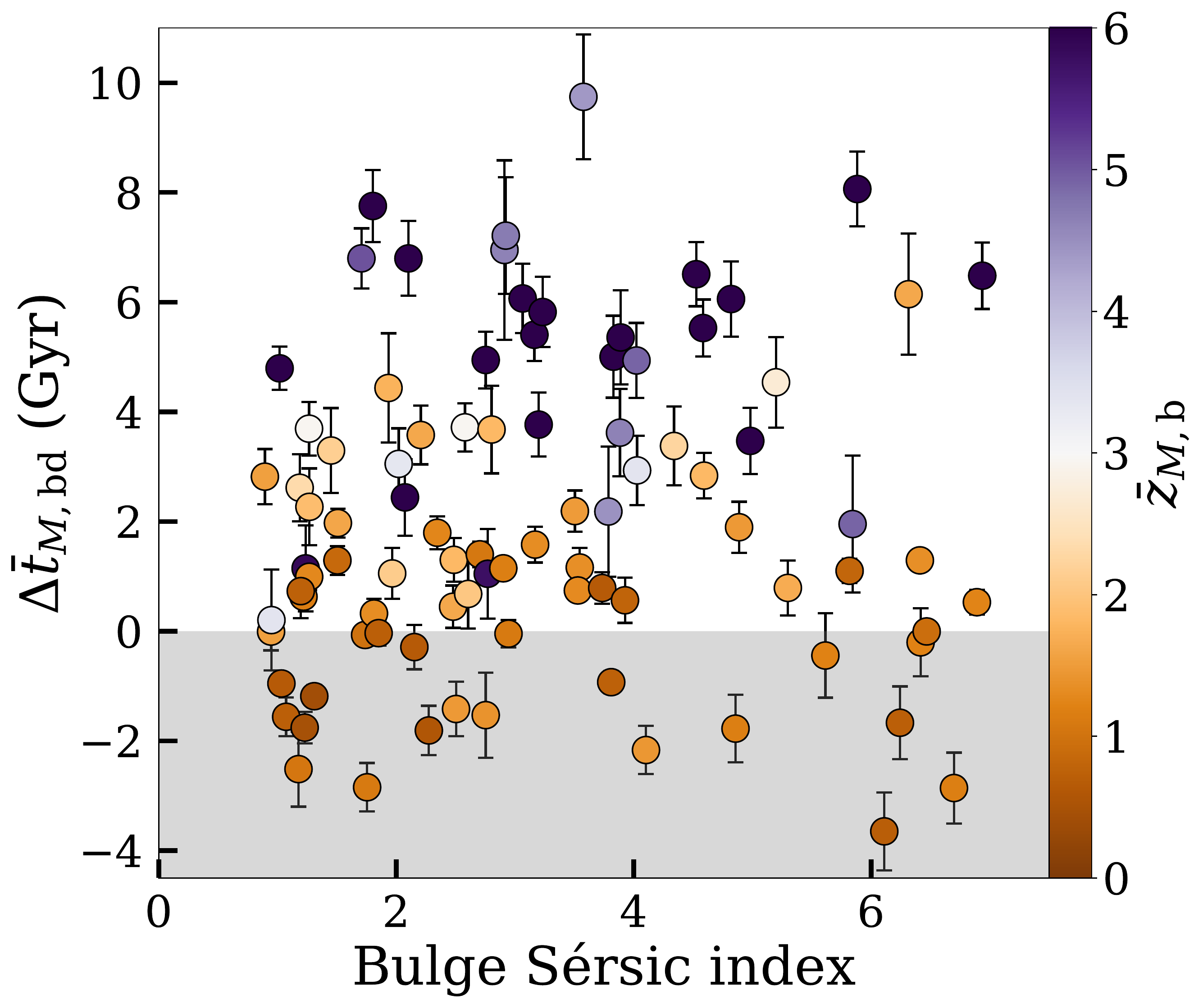}
\caption{Age difference of bulges and disks as a function of the bulge S\'ersic index,
color-coded according to the bulge mass-weighted formation redshift.
Errors are reported as 16th--84th percentile interval.
The gray shaded region marks \mbox{$\Delta \bar{t}_{M, \rm bd} < 0$}.
\label{fig:figure_9}}
\end{figure}

\subsubsection{Star Formation Timescale \label{sec:section_3_4_3}}

Disks of our massive galaxies present median 
$\tau_{\rm d} = 320^{+950}_{-110}$~Myr.
As highlighted in \citetalias{Costantin2021a}, bulges mostly formed on short timescales ($\tau \sim 200$~Myr)
and a slower mode of formation starts to be in place only for some second-wave bulges.
On the other hand, we find that 52\% of disks have $\tau_{\rm d} > 300$~Myr
and 38\% of disks have $\tau_{\rm d} > 500$~Myr.
The first disks in place ($\bar{z}_{M, \rm d} \sim 2-3$) formed on longer timescales
compared to the ones formed at lower redshift. Indeed, all disks formed 
at redshift $\bar{z}_{M, \rm d} > 1.7$ have $\tau_{\rm d} \gtrsim 800$~Myr.

Disks around first and second-wave bulges 
have median $\tau_{\rm d}= 360^{+610}_{-160}$~Myr and
$\tau_{\rm d} = 320^{+1110}_{-110}$~Myr, respectively.
A Kolmogorov-Smirnov test ($K = 0.11$, $p$-value $> 0.9$) suggests that 
we cannot reject the null hypothesis that these two distributions are similar.

We find that disks older than their bulges form on longer timescales
compared with the younger ones. Indeed, disks younger and older 
than bulges have $\tau_{\rm d} = 260^{+680}_{-60}$~Myr and
$\tau_{\rm d} = 900^{+550}_{-690}$~Myr, respectively.

\subsubsection{Mass Surface Density \label{sec:section_3_4_4}}

We parameterize the bulge and disk compactness looking at their mass surface density
$\Sigma_{1.5}$ (see Sect.~\ref{sec:section_2_3}).
As expected, we find that bulges and disks have very different mass surface densities:
bulges present median $\log(\Sigma_{1.5, \rm b}) = 10.4^{+0.5}_{-0.6}$~M$_{\odot}$~kpc$^{-1.5}$,
while disks have median $\log(\Sigma_{1.5, \rm d}) = 9.1^{+0.4}_{-0.6}$~M$_{\odot}$~kpc$^{-1.5}$.

We show in Fig.~\ref{fig:figure_8} the trend between 
$\bar{z}_{M, \rm d}$ and $\log(\Sigma_{1.5, \rm d})$.
There is a mild trend between the disk mass-weighted formation redshift and
their mass surface density (Spearman coefficient of $0.36$; $p$-value $< 0.01$),
i.e., older disks are more compact.

We already discussed in \citetalias{Costantin2021a} that first-wave bulges
display higher values of mass surface density than 
second-wave bulges, i.e., they are more compact 
($\log(\Sigma_{1.5, \rm b}) = 10.6^{+0.4}_{-0.4}$~M$_{\odot}$~kpc$^{-1.5}$ and
$\log(\Sigma_{1.5, \rm b}) = 10.2^{+0.5}_{-0.4}$~M$_{\odot}$~kpc$^{-1.5}$, respectively). 
This is a hint for characterizing their different formation 
mechanism. But, we find no differences of $\Sigma_{1.5, \rm d}$
between disks around first and second-wave bulges
($\log(\Sigma_{1.5, \rm d}) = 9.0^{+0.6}_{-0.5}$~M$_{\odot}$~kpc$^{-1.5}$
and $\log(\Sigma_{1.5, \rm d}) = 9.1^{+0.4}_{-0.7}$~M$_{\odot}$~kpc$^{-1.5}$, respectively).
Nonetheless, we find that the first structures to form are the more compact.
This is not only valid for galaxies, but also for each of their morphological components.

\begin{deluxetable*}{ccccccccc}
\tabletypesize{\small}
\tablecaption{Median physical properties of bulges and disks at redshift $0.14 < z \leq 1$. \label{tab:table_2}}
\tablehead{
\colhead{Type} & \colhead{$\log(M_{\star}$)} & \colhead{$\bar{t}_{M}$} & 
\colhead{$\bar{z}_{M}$} & \colhead{$z_{10}$} & \colhead{$z_{90}$} &
\colhead{$\tau$} & \colhead{$R_{\rm e}$} & \colhead{$\log(\Sigma_{1.5})$} \\
\colhead{} &\colhead{(M$_{\odot}$)} & \colhead{(Gyr)} & 
\colhead{} & \colhead{} & \colhead{} &
\colhead{(Myr)} & \colhead{(kpc)} & \colhead{(M$_{\odot}$ kpc$^{-1.5}$)} 
}
\decimalcolnumbers
\startdata
\textbf{bulges}						&	$10.5^{+0.3}_{-0.5}$		&	$2.7^{+3.9}_{-1.6}$	&	$1.6^{+4.6}_{-0.7}$	&	$2.0^{+6.8}_{-1.0}$	&	$1.3^{+3.4}_{-0.6}$	&	$210^{+480}_{-10}$		&	$1.0^{+0.9}_{-0.4}$	&	$10.4^{+0.5}_{-0.6}$	\\
\\
first-wave bulges					&	$10.8^{+0.2}_{-0.3}$		&	$6.5^{+1.5}_{-1.4}$	&	$6.2^{+1.5}_{-1.7}$	&	$8.8^{+3.9}_{-2.9}$	&	$4.7^{+0.7}_{-1.2}$	&	$200^{+20}_{-10}$		&	$1.3^{+0.8}_{-0.6}$	&	$10.6^{+0.4}_{-0.4}$	\\
second-wave bulges					&	$10.3^{+0.4}_{-0.5}$		&	$1.7^{+2.0}_{-0.8}$	&	$1.3^{+0.6}_{-0.6}$	&	$1.4^{+1.9}_{-0.6}$	&	$1.1^{+0.4}_{-0.4}$	&	$210^{+790}_{-10}$		&	$1.0^{+0.8}_{-0.4}$	&	$10.2^{+0.5}_{-0.4}$	\\
\\
bulges older than disks				&	$10.6^{+0.3}_{-0.5}$		&	$4.5^{+2.7}_{-2.8}$	&	$2.3^{+4.5}_{-1.0}$	&	$4.0^{+6.2}_{-2.6}$	&	$1.7^{+3.3}_{-0.6}$	&	$210^{+480}_{-10}$		&	$1.1^{+0.9}_{-0.4}$	&	$10.5^{+0.5}_{-0.6}$	\\
bulges coeval of disks				&	$10.4^{+0.1}_{-0.2}$		&	$1.4^{+1.9}_{-0.6}$	&	$1.1^{+0.4}_{-0.4}$	&	$1.2^{+0.4}_{-0.4}$	&	$1.0^{+0.3}_{-0.4}$	&	$230^{+80}_{-20}$		&	$1.0^{+0.4}_{-0.2}$	&	$10.3^{+0.3}_{-0.4}$	\\
bulges younger than disks				&	$10.1^{+0.2}_{-0.5}$		&	$1.1^{+0.4}_{-1.0}$	&	$0.8^{+0.5}_{-0.2}$	&	$0.9^{+0.6}_{-0.2}$	&	$0.8^{+0.4}_{-0.3}$	&	$200^{+510}_{-10}$		&	$0.7^{+0.7}_{-0.3}$	&	$10.1^{+0.5}_{-0.3}$	\\
\\
\textbf{disks}						&	$10.2^{+0.4}_{-0.6}$		&	$1.2^{+1.6}_{-0.9}$	&	$1.0^{+0.6}_{-0.3}$	&	$1.0^{+1.7}_{-0.4}$	&	$0.9^{+0.3}_{-0.3}$	&	$320^{+950}_{-110}$		&	$5.4^{+3.4}_{-2.0}$	&	$9.1^{+0.4}_{-0.6}$	\\
\\
disks around first-wave bulges			&	$10.3^{+0.3}_{-0.7}$		&	$1.3^{+1.8}_{-0.9}$	&	$0.9^{+0.9}_{-0.4}$	&	$1.1^{+2.1}_{-0.5}$	&	$0.8^{+0.4}_{-0.3}$	&	$360^{+610}_{-160}$	&	$6.3^{+3.1}_{-2.5}$	&	$9.0^{+0.6}_{-0.5}$	\\
disks around second-wave bulges		&	$10.2^{+0.3}_{-0.5}$		&	$1.0^{+1.7}_{-0.8}$	&	$1.0^{+0.6}_{-0.3}$	&	$1.0^{+1.1}_{-0.2}$	&	$0.9^{+0.2}_{-0.3}$	&	$320^{+1110}_{-110}$	&	$4.8^{+2.8}_{-1.4}$	&	$9.1^{+0.4}_{-0.7}$	\\
\\
disks younger than bulges				&	$10.2^{+0.3}_{-0.6}$		&	$0.8^{+0.9}_{-0.6}$	&	$0.9^{+0.4}_{-0.3}$	&	$1.0^{+0.6}_{-0.3}$	&	$0.9^{+0.2}_{-0.3}$	&	$260^{+680}_{-60}$		&	$6.0^{+3.1}_{-2.4}$	&	$9.0^{+0.5}_{-0.5}$	\\
disks coeval of bulges				&	$10.4^{+0.3}_{-0.6}$		&	$1.6^{+2.1}_{-0.7}$	&	$1.2^{+0.3}_{-0.4}$	&	$1.6^{+1.8}_{-0.7}$	&	$0.9^{+0.2}_{-0.2}$	&	$750^{+2140}_{-340}$	&	$6.9^{+0.5}_{-3.4}$	&	$9.2^{+0.6}_{-0.8}$	\\
disks older than bulges				&	$10.3^{+0.3}_{-0.2}$		&	$2.9^{+0.9}_{-1.7}$	&	$1.5^{+1.0}_{-0.6}$	&	$2.8^{+2.5}_{-1.9}$	&	$0.9^{+0.4}_{-0.3}$	&	$900^{+550}_{-690}$	&	$3.6^{+1.9}_{-0.6}$	&	$9.4^{+0.4}_{-0.3}$	\\
\enddata
\tablecomments{Column (1): morphological component. Column (2): stellar mass. Column (3): mass-weighted age.
Column (4): mass-weighted formation redshift. Column (5): redshift when a component grows 10\% of its current mass. 
Column (6): redshift when a component grows 90\% of its current mass.
Column (7): timescale of exponentially declined SFH. 
Column (8): effective radius. Column (9): mass surface density. 
}
\end{deluxetable*}

\subsubsection{S\'ersic Index  \label{sec:section_3_4_5}}

We find that neither the mass-weighted formation redshift of bulges nor those
of disks correlates with the bulge S\'ersic index, having
Spearman coefficient of $0.12$ and $-0.01$, respectively ($p$-value $> 0.25$ and $0.90$, respectively).
Moreover, bulge and disk mass-weighted formation redshift do not correlate with the S\'ersic index of the 
galaxy either, having Spearman coefficient of $0.15$ and $0.17$, respectively 
($p$-value $> 0.15$ and $0.11$, respectively).
As shown in Fig.~\ref{fig:figure_9}, no correlation is found between the $\Delta \bar{t}_{M, \rm bd}$
and the bulge S\'ersic index, being the Spearman coefficient $\sim0.13$ 
($p$-value $> 0.21$).


\section{Discussion \label{sec:section_4}}

\begin{figure*}[t!]
\centering
\includegraphics[scale=0.5]{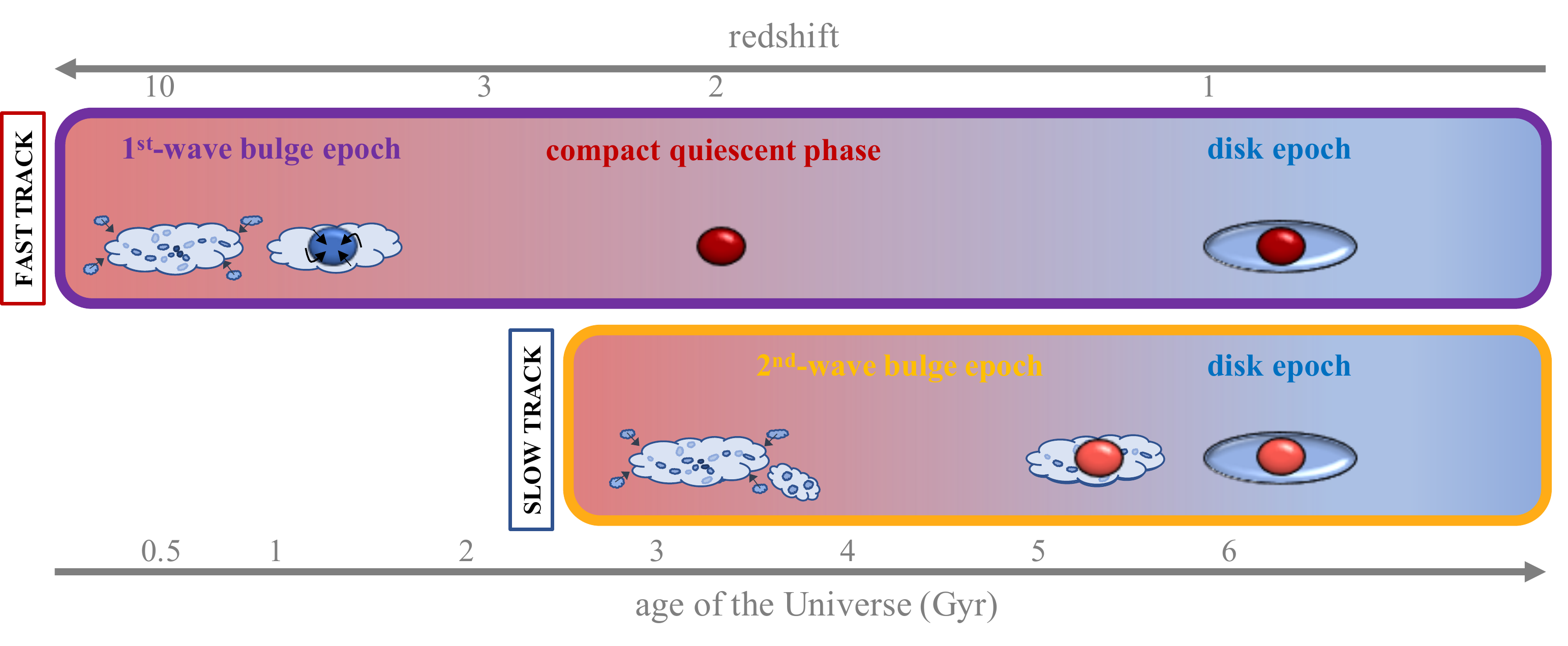}
\caption{Illustration of the proposed scenario for the formation and morphological evolution of
massive disk galaxies at redshift $0.14 < z \leq 1$.
This cartoon complement the picture detailed in \citetalias{Costantin2021a} (Fig.~13), 
adding the information about the disk growth.
The upper panel shows the evolution of fast-track systems. 
These galaxies build a compact spheroid at high redshift through an intense episode of star formation (first-wave bulge),
evolve rapidly through a blue and red-nugget phase ($z\sim1.5-3$), and grow an extended stellar disk by redshift $z\sim1$.
The lower panel shows the evolution of slow-track systems. 
In these galaxies there is a high level of (slow) co-evolution between the spheroidal 
(second-wave bulge; $z<3$) and disk component ($z\sim1$) 
and probably no compact quiescent phase would be observed, i.e., a relatively prominent star-forming disk is always present.
In the cartoon, the difference in age between first and second-wave bulges is marked by darker to lighter red colors
in the disk galaxies sketched at redshift $z<1$.
\label{fig:figure_10}}
\end{figure*}

In this Section we discuss the main results of our analysis,
summarizing our findings about bulge and disk formation 
(presented in \citetalias{Costantin2021a} and this paper).
To facilitate the comparison, 
we briefly outline all the physical properties measured for our bulges and disks (in Table~\ref{tab:table_2}), 
dividing systems between first and second wave as well as separating galaxies 
in those with bulges older, coeval, or younger than their disks.
Moreover, we qualitatively illustrate in Fig.~\ref{fig:figure_10} the proposed scenario for the formation and 
morphological evolution of massive disk galaxies at redshift $0.14 < z \leq1$,
complementing the picture provided in \citetalias{Costantin2021a} (Fig.~13),
adding the information about the disk epoch.

Galaxies are complex systems, and their integrated properties
are hiding the complexity of their evolution.
Indeed, we infer from our results that 
we are looking at an evolutionary sequence in which
massive disk galaxies shape their morphology,
growing their extended stellar disks around a centrally concentrated spheroid.
The residual and continuous gas accretion enables the central spheroid to slowly develop
a stellar envelope, allowing the incoming gas to retain angular momentum and
resulting in a more extended and disky system by $z\lesssim1$
\citep[see also][]{PerezGonzalez2008, Buitrago2013}.
In particular, we find that disks form efficiently when bulges grow 
inefficiently and viceversa (Figs.~\ref{fig:figure_4} and \ref{fig:figure_5}).
Indeed, we report a delay in the time of formation of the two components.
This is consistent with recent cosmological simulations (e.g., IllustrisTNG),
showing that bulges form efficiently at early cosmic times
and assembled most of their mass by the time galaxies stop forming stars \citep{Tacchella2019}.

Our main result is that disks consistently form at $z \sim 1$ around first and second-wave bulges.
This evidence may suggest that disks at these 
redshifts are not destroyed, while they might be at higher redshift ($z \gtrsim 3-5$).
The reasons for this might be the enhanced merger activities and the (larger) more turbulent gas fractions.
Thus, primordial (dynamically unstable) gas-riched disks have a different fate than 
extended thin (dynamically stable) disks which grow at $z\lesssim1$.
Furthermore, also first and second-wave bulges show very different paths of evolution,
being the first one the end products of compaction
events at very high redshift (\citetalias{Costantin2021a}).

In the two-phase scenario proposed for early-type galaxies \citep{Oser2010},
a compact progenitor rapidly built at high redshift \citep{Dekel2009,Zolotov2015} and
slowly grew in size through non-dissipational processes (e.g., dry minor mergers)
until resulting in an elliptical galaxy in the local universe \citep{Naab2009, Huang2013a, Huang2013b}.
In this picture, we argue that also massive disk galaxies could play a crucial role,
hosting a compact core (first-wave bulge) which went through a blue and red-nugget 
phase and grew an extended stellar disk at later times. 
Moreover, some of second-wave bulges are as compact as
first-wave bulges and formed before their disks. They could be interpreted as a later wave of systems 
going through a red-nugget phase at $z<3$,
but keeping in mind that these galaxies experienced a higher degree of co-evolution
between the spheroidal and disk component (see Sect.~\ref{sec:section_3_3_2}).
This could allow to extend the two-phase paradigm to late-type galaxies,
as already proposed by recent studies \citep{Graham2013, delaRosa2016, Costantin2020}.

We remark that only a proper decoupling of the two components
allows us to characterize the instants where these galaxies form
and the physical processes responsible for their evolution.
In this picture, the results of our work point toward a scenario where the majority of 
massive disk galaxies assemble inside-out. 
In $\sim75\%$ of our galaxies
the central bulge forms half of its mass earlier than the disk component,
which takes on average $\sim 1.5$~Gyr to develop around the bulge. 
In particular, for $\sim 35\%$ of the systems the bulge takes more than 3~Gyr to develop an extended disk.
This result is consistent with predictions from EAGLE cosmological simulations
\citep{Schaye2015, Crain2015} showing that on average
disk-dominated galaxies more massive than $10^{10}$~M$_{\odot}$ 
have inner regions older than the outer part ($\sim 1$~Gyr at $z=0.5$), 
compatible with an inside-out formation \citep{Pfeffer2022}.

The few disks (17\%) older than their bulges are characterized 
by longer timescales with respect to disks younger than their bulges.
Considering that some disks start to form as early as some bulges, 
but they assemble their mass slower, this suggests a high degree of co-evolution
between the two components. Disk material is probably accreted into the central
region of the galaxy, funneling gas and highly increasing the star formation efficiency of the spheroidal component.
However, since bulges build up fast, as time passes we see residual star formation
in the disk component simultaneously with a reduction of the star formation activity in the central region of the galaxy.

Focusing on the first-assembled systems ($z\sim10$),
we find that they grow an extended stellar disk from a spheroidal-like system (first-wave bulge).
In this scenario, cosmological simulations show that it is possible to form
extended star-forming disks around red nuggets after compaction \citep{Zolotov2015, Dekel2020}.
Given that first-wave bulges are very massive and compact, and considering the mass-weighted age differences 
between the bulge and the disk in these galaxies, our results point to a 
morphological quenching and a stabilization of the galaxies 
which prevents further star formation in the primordial disk for several Gyrs.
In these galaxies, it is possible that the compaction phase 
of first-wave bulges consumes the cold gas in the outskirts 
(or quickly destroy the primordial disk)
and inhibits the star formation 
for as long as $\sim8$~Gyrs (see Fig.~\ref{fig:figure_6}).
Even if some disks start to build at redshift $z\sim5$, they grow slowly
($\tau_{\rm d} \sim 1$~Gyr)
due to violent disk instabilities and clump migration,
which result in a low cold gas fraction available in the outskirts of these systems
or with a lower efficiency of star formation.
Similarly, the star formation in disks around second-wave bulges may be lower 
because the recent growth of the bulge has depleted the cold gas in the outskirts of the galaxy.
In the New Horizon cosmological simulations, the driver of the bulge growth in disk-dominated galaxies
is the increase of perturbed disk stars at early cosmic time until $z\sim1.5$ \citep{Park2019},
which results into a lower efficiency for the formation of stars in the disk.
Again, this migration of stars from an unstable primordial disk into a compact spheroid
agrees with the compaction events which are responsible to form red-nugget systems \citep{Ceverino2015, Zolotov2015}.

Several studies described the inside-out growth of massive galaxies
by studying the radial gradient of their stellar mass surface density 
\citep[e.g., ][]{vanDokkum2010, Patel2013}. By linking progenitors and descendants of these galaxies, 
these works show that massive galaxies have assembled their extended stellar 
halos around compact and dense cores.
Similarly, the observed size evolution of star-forming galaxies
was interpreted as a different distribution of their stellar populations,
where the youngest stars have a more extended distribution than the older stars \citep[e.g.,][]{Williams2010}.
Our analysis adds a fundamental piece of information to these studies:
we morphologically separate the bulge and disk 
components and derive their stellar population properties,
providing clues about the assembly time for each component separately.

Interestingly, we find that not only galaxies form
in a downsizing fashion, but also each of their morphological components does.
As shown in Figs.~\ref{fig:figure_5} and \ref{fig:figure_6}, 
galaxies, bulges, and disks form their stars later as they are less massive.
Interestingly, more massive system are not only older, but also more compact.

We find that the inside-out growth of massive disk galaxies strongly depends on 
the bulge mass (see Fig.~\ref{fig:figure_6}), which actually drives the morphological evolution of our galaxies.
Our results complement and extend (up to redshift $z\sim1$) the results obtained 
by \citet{MendezAbreu2021} studying the evolution
of bulges and disks in local galaxies within the CALIFA survey \citep{Sanchez2012}.
The properties of bulges drive the future evolution of the galaxy as a whole, 
while disks have properties being set up by those of the galaxy, but not affecting them.


\section{Conclusions \label{sec:section_5}}

In this work we have investigated how massive disk galaxies
shape their morphology across cosmic time.
We studied a sample of galaxies from the SHARDS spectro-photometric survey in GOODS-N,
which we photometrically modeled as a central bulge and an extended stellar disk.
Thanks to the SHARDS data and also counting with the exquisite morphological 
information provided by HST/CANDELS data, we retrieved the SEDs of each bulge 
and disk in those galaxies with a spectral resolution $R\sim50$. 
The spectral resolution and depth of the SHARDS data allowed us to characterize their
individual SFHs fitting the SEDs to stellar population synthesis models.

We find that the majority ($\sim85\%$) of massive disk galaxies
grows inside-out. The peak in the formation of bulges in massive galaxies
at $0.14 < z \leq 1$ occurred at $\bar{z}_{M, \rm b}=1.6$, with a first-wave population
building half of their mass as early as 0.9~Gyr after the Big Bang ($\bar{z}_{M, \rm b}=6.2$) 
and a second wave peaking 3.8~Gyr later ($\bar{z}_{M, \rm b}=1.3$).
In contrast, the disks in these galaxies typically formed at $\bar{z}_{M, \rm d}=1$.
The bulges formed in a first wave at earlier cosmic times took longer
(5.2 Gyr) to grow a disk than the bulges in the second wave (which took 0.7 Gyr), 
many of the latter still showing significant star formation activity in their disks.

There are a few disks ($\sim15\%$) that started to assemble as early as first-wave bulges ($z\gtrsim3$),
but they grow on longer timescale ($\tau_{\rm d} \gtrsim 1$~Gyr) 
compared to bulges ($\tau_{\rm b} \lesssim 300$~Myr).
Similarly, second-wave disks also assemble on longer timescales, 
suggesting a higher degree of co-evolution between the bulge and 
disk components in galaxies at redshift $z\lesssim1.5$ compared to higher-redshift ones.
The average rate of star formation for disks which develop 
around first-wave bulges could be $2-3$ times more intense 
than the one of disks around second-wave bulges.

Importantly, we find that not only galaxies grow in a downsizing fashion, 
but also each of their morphological components does:
both older disks and older bulges are more massive than younger stellar structures.
In addition, the oldest and most massive bulges and disks are also the most compact ones.

Accordingly with the latest results on the formation of nearby disk galaxies from \citet{MendezAbreu2021}, 
we find that the mass of the bulge regulates the timing of the growth of the extended stellar disk.
In particular, galaxies hosting the more compact (first-wave) bulges took longer to acquire their disks
than galaxies containing second-wave bulges.
But, we do not find distinct physical properties (e.g., mass, star formation timescale, mass surface density, and S\'ersic index) 
for the disks in both types of galaxies.
Thus, since disks consistently form at $z\sim1$ around first and second-wave bulges,
we conclude the only way to distinguish the formation mechanisms of these galaxies
is to disentangle the SFHs of their disks from the ones of bulges.
Indeed, the mechanisms which drive the formation of massive disk galaxies 
left imprints on the observed properties of their first and second-wave bulges.

\acknowledgments

We would like to thank the anonymous referee for improving the content of the manuscript. 
We are grateful to Ignacio Trujillo and Christoper J.~Conselice for the useful discussions
and comments.
LC wishes to thank Cristina Cabello for the support provided while 
this project was devised and Michele Perna for the fruitful discussions.
LC acknowledges financial support from Comunidad de Madrid under 
Atracci\'on de Talento grant 2018-T2/TIC-11612.
LC and PGPG acknowledge support from
Spanish Ministerio de Ciencia, Innovaci\'on y Universidades through grant PGC2018-093499-B-I00.
JMA acknowledges the support of the Viera y Clavijo Senior program funded by ACIISI and ULL.
DC is a Ramon-Cajal Researcher and is supported by the Ministerio de Ciencia, 
Innovaci\'on y Universidades (MICIU/FEDER) under research grant PGC2018-094975-C21.
This work has made use of the Rainbow Cosmological Surveys Database, 
which is operated by the Centro de Astrobiolog\'ia (CAB/INTA), 
partnered with the University of California Observatories at Santa Cruz (UCO/Lick,UCSC).


\appendix

\section{Mass-weighted Ages \label{app:A}}

\begin{figure*}[t!]
\centering
\includegraphics[scale=0.4, trim=0cm 0cm 0cm 0cm , clip=true]{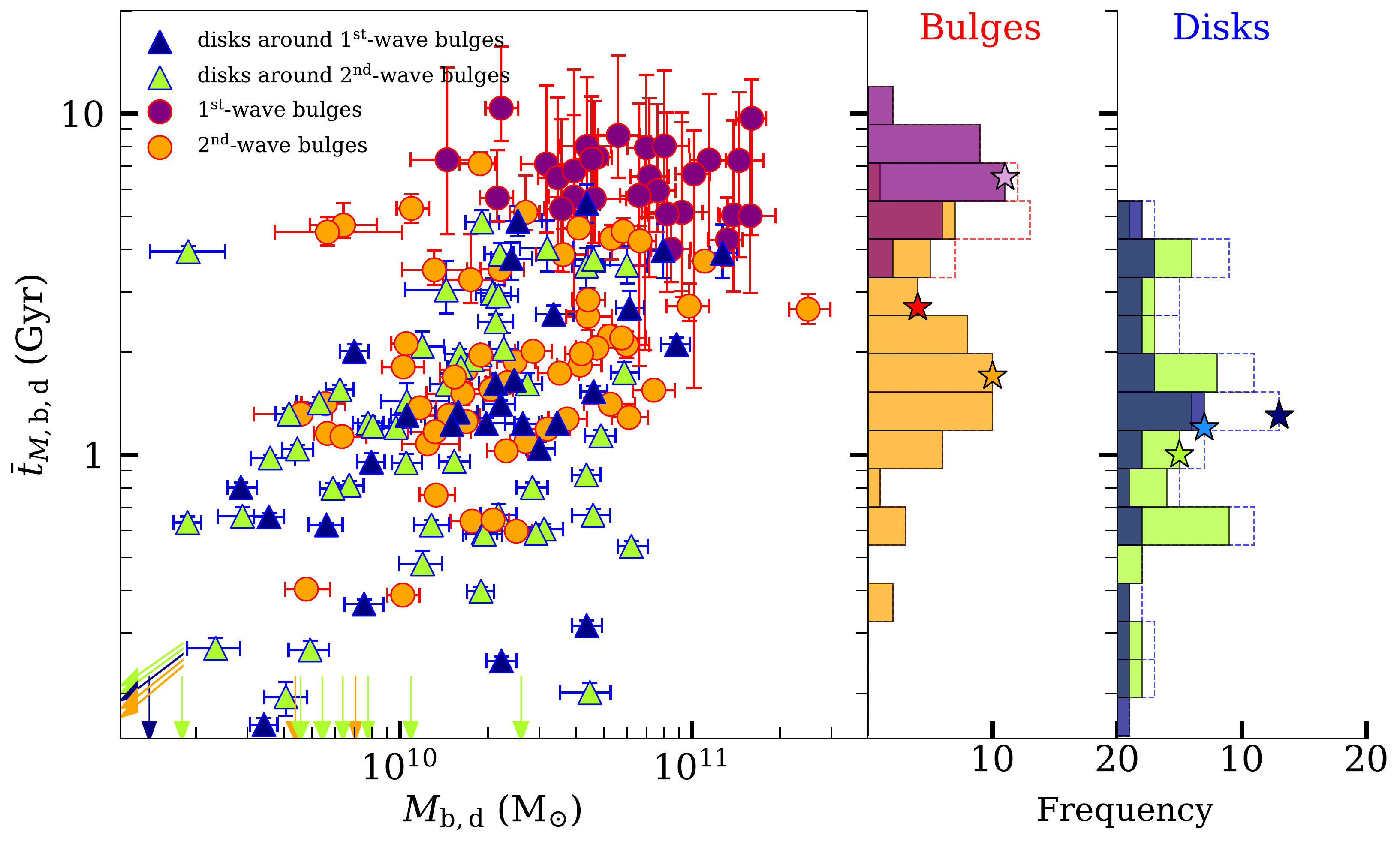}
\caption{Mass-weighted stellar ages of bulges (dots) and disks (triangles) as a function
of their stellar mass. Purple and blue symbols stand for first-wave bulges ($\bar{z}_{M, \rm b} > 3$) and disks around them,
while orange and green symbols represent second-wave bulges ($\bar{z}_{M, \rm b} < 3$) and disks around them, respectively.
Errors are reported as 16th--84th percentile interval.
Arrows mark upper limits for bulges and disks 
with mass-weighted ages $\bar{t}_{\rm M} < 150$~Myr, respectively.
Purple and orange histograms stand for first and second-wave bulges,
while the histogram with red dashed contour stands for the entire bulge population.
Blue and green histograms stand for disks around first and second-wave bulges,
while the histogram with light blue dashed contour stands for the entire disk population.
The median values of each distribution are marked with stars.
\label{fig:figure_A}}
\end{figure*}

In this Section we directly compare the age distributions of each bulge and disk,
explicitly showing what we claimed in Sects.~\ref{sec:section_3_1} and \ref{sec:section_3_2}: 
on average bulges are older than disks (Fig.~\ref{fig:figure_A}).
Bulges have median mass-weighted ages $\bar{t}_{M, \rm b} = 2.7^{+3.9}_{-1.6}$~Gyr,
while disks present $\bar{t}_{M, \rm d} = 1.2^{+1.6}_{-0.9}$~Gyr.
This means that, on average, spheroidal galaxies take $\sim1.5$~Gyr 
in acquiring a stellar disk (see Sect.~\ref{sec:section_3_3_2}).
In \citetalias{Costantin2021a} we found that 48\% of our bulges have $\bar{t}_{M, \rm b} > 3$~Gyr.
In contrast, only 14\% of disks have $\bar{t}_{M, \rm d} > 3$~Gyr.
7\% of bulges and 23\% of disks 
present significant ongoing star formation and/or assembly activity
($\bar{t}_{M} < 500$~Myr) at the redshift of observation.

We find that first-wave bulges have $\bar{t}_{M, \rm b} = 6.5^{+1.5}_{-1.4}$~Gyr while the disks around them present
median $\bar{t}_{M, \rm d} = 1.3^{+1.8}_{-0.9}$~Gyr. On the other hand, second-wave bulges 
have median $\bar{t}_{M, \rm b} = 1.7^{+2.0}_{-0.8}$~Gyr while the disks around them are
$1.0^{+1.7}_{-0.8}$~Gyr old.

\begin{thebibliography}{}
\expandafter\ifx\csname natexlab\endcsname\relax\def\natexlab#1{#1}\fi
\providecommand{\url}[1]{\href{#1}{#1}}
\providecommand{\dodoi}[1]{doi:~\href{http://doi.org/#1}{\nolinkurl{#1}}}
\providecommand{\doeprint}[1]{\href{http://ascl.net/#1}{\nolinkurl{http://ascl.net/#1}}}
\providecommand{\doarXiv}[1]{\href{https://arxiv.org/abs/#1}{\nolinkurl{https://arxiv.org/abs/#1}}}

\bibitem[{{Athanassoula}(2005)}]{Athanassoula2005}
{Athanassoula}, E. 2005, \mnras, 358, 1477,
  \dodoi{10.1111/j.1365-2966.2005.08872.x}

\bibitem[{{Barro} {et~al.}(2013){Barro}, {Faber}, {P{\'e}rez-Gonz{\'a}lez},
  {Koo}, {Williams}, {Kocevski}, {Trump}, {Mozena}, {McGrath}, {van der Wel},
  {Wuyts}, {Bell}, {Croton}, {Ceverino}, {Dekel}, {Ashby}, {Cheung},
  {Ferguson}, {Fontana}, {Fang}, {Giavalisco}, {Grogin}, {Guo}, {Hathi},
  {Hopkins}, {Huang}, {Koekemoer}, {Kartaltepe}, {Lee}, {Newman}, {Porter},
  {Primack}, {Ryan}, {Rosario}, {Somerville}, {Salvato}, \& {Hsu}}]{Barro2013}
{Barro}, G., {Faber}, S.~M., {P{\'e}rez-Gonz{\'a}lez}, P.~G., {et~al.} 2013,
  \apj, 765, 104, \dodoi{10.1088/0004-637X/765/2/104}

\bibitem[{{Barro} {et~al.}(2019){Barro}, {P{\'e}rez-Gonz{\'a}lez}, {Cava},
  {Brammer}, {Pandya}, {Eliche Moral}, {Esquej}, {Dom{\'\i}nguez-S{\'a}nchez},
  {Alcalde Pampliega}, {Guo}, {Koekemoer}, {Trump}, {Ashby}, {Cardiel},
  {Castellano}, {Conselice}, {Dickinson}, {Dolch}, {Donley}, {Espino Briones},
  {Faber}, {Fazio}, {Ferguson}, {Finkelstein}, {Fontana}, {Galametz},
  {Gardner}, {Gawiser}, {Giavalisco}, {Grazian}, {Grogin}, {Hathi}, {Hemmati},
  {Hern{\'a}n-Caballero}, {Kocevski}, {Koo}, {Kodra}, {Lee}, {Lin}, {Lucas},
  {Mobasher}, {McGrath}, {Nandra}, {Nayyeri}, {Newman}, {Pforr}, {Peth},
  {Rafelski}, {Rodr{\'\i}guez-Munoz}, {Salvato}, {Stefanon}, {van der Wel},
  {Willner}, {Wiklind}, \& {Wuyts}}]{Barro2019}
{Barro}, G., {P{\'e}rez-Gonz{\'a}lez}, P.~G., {Cava}, A., {et~al.} 2019, \apjs,
  243, 22, \dodoi{10.3847/1538-4365/ab23f2}

\bibitem[{{Belli} {et~al.}(2019){Belli}, {Newman}, \& {Ellis}}]{Belli2019}
{Belli}, S., {Newman}, A.~B., \& {Ellis}, R.~S. 2019, \apj, 874, 17,
  \dodoi{10.3847/1538-4357/ab07af}

\bibitem[{{Bournaud} {et~al.}(2007){Bournaud}, {Elmegreen}, \&
  {Elmegreen}}]{Bournaud2007}
{Bournaud}, F., {Elmegreen}, B.~G., \& {Elmegreen}, D.~M. 2007, \apj, 670, 237,
  \dodoi{10.1086/522077}

\bibitem[{{Brook} {et~al.}(2006){Brook}, {Kawata}, {Martel}, {Gibson}, \&
  {Bailin}}]{Brook2006}
{Brook}, C.~B., {Kawata}, D., {Martel}, H., {Gibson}, B.~K., \& {Bailin}, J.
  2006, \apj, 639, 126, \dodoi{10.1086/499154}

\bibitem[{{Bruce} {et~al.}(2014){Bruce}, {Dunlop}, {McLure}, {Cirasuolo},
  {Buitrago}, {Bowler}, {Targett}, {Bell}, {McIntosh}, {Dekel}, {Faber},
  {Ferguson}, {Grogin}, {Hartley}, {Kocevski}, {Koekemoer}, {Koo}, \&
  {McGrath}}]{Bruce2014}
{Bruce}, V.~A., {Dunlop}, J.~S., {McLure}, R.~J., {et~al.} 2014, \mnras, 444,
  1660, \dodoi{10.1093/mnras/stu1537}

\bibitem[{{Bruzual} \& {Charlot}(2003)}]{Bruzual2003}
{Bruzual}, G., \& {Charlot}, S. 2003, \mnras, 344, 1000,
  \dodoi{10.1046/j.1365-8711.2003.06897.x}

\bibitem[{{Buitrago} {et~al.}(2013){Buitrago}, {Trujillo}, {Conselice}, \&
  {H{\"a}u{\ss}ler}}]{Buitrago2013}
{Buitrago}, F., {Trujillo}, I., {Conselice}, C.~J., \& {H{\"a}u{\ss}ler}, B.
  2013, \mnras, 428, 1460, \dodoi{10.1093/mnras/sts124}

\bibitem[{{Calzetti} {et~al.}(2000){Calzetti}, {Armus}, {Bohlin}, {Kinney},
  {Koornneef}, \& {Storchi-Bergmann}}]{Calzetti2000}
{Calzetti}, D., {Armus}, L., {Bohlin}, R.~C., {et~al.} 2000, \apj, 533, 682,
  \dodoi{10.1086/308692}

\bibitem[{{Carnall} {et~al.}(2019){Carnall}, {McLure}, {Dunlop}, {Cullen},
  {McLeod}, {Wild}, {Johnson}, {Appleby}, {Dav{\'e}}, {Amorin}, {Bolzonella},
  {Castellano}, {Cimatti}, {Cucciati}, {Gargiulo}, {Garilli}, {Marchi},
  {Pentericci}, {Pozzetti}, {Schreiber}, {Talia}, \& {Zamorani}}]{Carnall2019}
{Carnall}, A.~C., {McLure}, R.~J., {Dunlop}, J.~S., {et~al.} 2019, \mnras, 490,
  417, \dodoi{10.1093/mnras/stz2544}

\bibitem[{{Ceverino} {et~al.}(2015){Ceverino}, {Dekel}, {Tweed}, \&
  {Primack}}]{Ceverino2015}
{Ceverino}, D., {Dekel}, A., {Tweed}, D., \& {Primack}, J. 2015, \mnras, 447,
  3291, \dodoi{10.1093/mnras/stu2694}

\bibitem[{{Ceverino} {et~al.}(2018){Ceverino}, {Klessen}, \&
  {Glover}}]{Ceverino2018}
{Ceverino}, D., {Klessen}, R.~S., \& {Glover}, S. C.~O. 2018, \mnras, 480,
  4842, \dodoi{10.1093/mnras/sty2124}

\bibitem[{{Chabrier}(2003)}]{Chabrier2003}
{Chabrier}, G. 2003, \pasp, 115, 763, \dodoi{10.1086/376392}

\bibitem[{{Clauwens} {et~al.}(2018){Clauwens}, {Schaye}, {Franx}, \&
  {Bower}}]{Clauwens2018}
{Clauwens}, B., {Schaye}, J., {Franx}, M., \& {Bower}, R.~G. 2018, \mnras, 478,
  3994, \dodoi{10.1093/mnras/sty1229}

\bibitem[{{Cole} {et~al.}(2000){Cole}, {Lacey}, {Baugh}, \& {Frenk}}]{Cole2000}
{Cole}, S., {Lacey}, C.~G., {Baugh}, C.~M., \& {Frenk}, C.~S. 2000, \mnras,
  319, 168, \dodoi{10.1046/j.1365-8711.2000.03879.x}

\bibitem[{{Costantin} {et~al.}(2018{\natexlab{a}}){Costantin}, {Corsini},
  {M{\'e}ndez-Abreu}, {Morelli}, {Dalla Bont{\`a}}, \&
  {Pizzella}}]{Costantin2018b}
{Costantin}, L., {Corsini}, E.~M., {M{\'e}ndez-Abreu}, J., {et~al.}
  2018{\natexlab{a}}, \mnras, 481, 3623, \dodoi{10.1093/mnras/sty1754}

\bibitem[{{Costantin} {et~al.}(2018{\natexlab{b}}){Costantin},
  {M{\'e}ndez-Abreu}, {Corsini}, {Eliche-Moral}, {Tapia}, {Morelli}, {Dalla
  Bont{\`a}}, \& {Pizzella}}]{Costantin2018a}
{Costantin}, L., {M{\'e}ndez-Abreu}, J., {Corsini}, E.~M., {et~al.}
  2018{\natexlab{b}}, \aap, 609, A132, \dodoi{10.1051/0004-6361/201731823}

\bibitem[{{Costantin} {et~al.}(2017){Costantin}, {M{\'e}ndez-Abreu}, {Corsini},
  {Morelli}, {Aguerri}, {Dalla Bont{\`a}}, \& {Pizzella}}]{Costantin2017}
---. 2017, \aap, 601, A84, \dodoi{10.1051/0004-6361/201630302}

\bibitem[{{Costantin} {et~al.}(2020){Costantin}, {M{\'e}ndez-Abreu}, {Corsini},
  {Morelli}, {de Lorenzo-C{\'a}ceres}, {Pagotto}, {Cuomo}, {Aguerri}, \&
  {Rubino}}]{Costantin2020}
---. 2020, \apjl, 889, L3, \dodoi{10.3847/2041-8213/ab6459}

\bibitem[{{Costantin} {et~al.}(2021){Costantin}, {P{\'e}rez-Gonz{\'a}lez},
  {M{\'e}ndez-Abreu}, {Huertas-Company}, {Dimauro}, {Alcalde-Pampliega},
  {Buitrago}, {Ceverino}, {Daddi}, {Dom{\'\i}nguez-S{\'a}nchez},
  {Espino-Briones}, {Hern{\'a}n-Caballero}, {Koekemoer}, \&
  {Rodighiero}}]{Costantin2021a}
{Costantin}, L., {P{\'e}rez-Gonz{\'a}lez}, P.~G., {M{\'e}ndez-Abreu}, J.,
  {et~al.} 2021, \apj, 913, 125, \dodoi{10.3847/1538-4357/abef72}

\bibitem[{{Crain} {et~al.}(2015){Crain}, {Schaye}, {Bower}, {Furlong},
  {Schaller}, {Theuns}, {Dalla Vecchia}, {Frenk}, {McCarthy}, {Helly},
  {Jenkins}, {Rosas-Guevara}, {White}, \& {Trayford}}]{Crain2015}
{Crain}, R.~A., {Schaye}, J., {Bower}, R.~G., {et~al.} 2015, \mnras, 450, 1937,
  \dodoi{10.1093/mnras/stv725}

\bibitem[{{Dalcanton} {et~al.}(1997){Dalcanton}, {Spergel}, \&
  {Summers}}]{Dalcanton1997}
{Dalcanton}, J.~J., {Spergel}, D.~N., \& {Summers}, F.~J. 1997, \apj, 482, 659,
  \dodoi{10.1086/304182}

\bibitem[{{Damjanov} {et~al.}(2009){Damjanov}, {McCarthy}, {Abraham},
  {Glazebrook}, {Yan}, {Mentuch}, {Le Borgne}, {Savaglio}, {Crampton},
  {Murowinski}, {Juneau}, {Carlberg}, {J{\o}rgensen}, {Roth}, {Chen}, \&
  {Marzke}}]{Damjanov2009}
{Damjanov}, I., {McCarthy}, P.~J., {Abraham}, R.~G., {et~al.} 2009, \apj, 695,
  101, \dodoi{10.1088/0004-637X/695/1/101}

\bibitem[{{de la Rosa} {et~al.}(2016){de la Rosa}, {La Barbera}, {Ferreras},
  {S{\'a}nchez Almeida}, {Dalla Vecchia}, {Mart{\'\i}nez-Valpuesta}, \&
  {Stringer}}]{delaRosa2016}
{de la Rosa}, I.~G., {La Barbera}, F., {Ferreras}, I., {et~al.} 2016, \mnras,
  457, 1916, \dodoi{10.1093/mnras/stw130}

\bibitem[{{de Lorenzo-C{\'a}ceres} {et~al.}(2019{\natexlab{a}}){de
  Lorenzo-C{\'a}ceres}, {M{\'e}ndez-Abreu}, {Thorne}, \&
  {Costantin}}]{deLorenzoCacered2019a}
{de Lorenzo-C{\'a}ceres}, A., {M{\'e}ndez-Abreu}, J., {Thorne}, B., \&
  {Costantin}, L. 2019{\natexlab{a}}, \mnras, 484, 665,
  \dodoi{10.1093/mnras/sty3520}

\bibitem[{{de Lorenzo-C{\'a}ceres} {et~al.}(2019{\natexlab{b}}){de
  Lorenzo-C{\'a}ceres}, {S{\'a}nchez-Bl{\'a}zquez}, {M{\'e}ndez-Abreu},
  {Gadotti}, {Falc{\'o}n-Barroso}, {Mart{\'\i}nez-Valpuesta}, {Coelho},
  {Fragkoudi}, {Husemann}, {Leaman}, {P{\'e}rez}, {Querejeta}, {Seidel}, \&
  {van de Ven}}]{deLorenzoCacered2019b}
{de Lorenzo-C{\'a}ceres}, A., {S{\'a}nchez-Bl{\'a}zquez}, P.,
  {M{\'e}ndez-Abreu}, J., {et~al.} 2019{\natexlab{b}}, \mnras, 484, 5296,
  \dodoi{10.1093/mnras/stz221}

\bibitem[{{Dekel} \& {Burkert}(2014)}]{Dekel2014}
{Dekel}, A., \& {Burkert}, A. 2014, \mnras, 438, 1870,
  \dodoi{10.1093/mnras/stt2331}

\bibitem[{{Dekel} {et~al.}(2009){Dekel}, {Sari}, \& {Ceverino}}]{Dekel2009}
{Dekel}, A., {Sari}, R., \& {Ceverino}, D. 2009, \apj, 703, 785,
  \dodoi{10.1088/0004-637X/703/1/785}

\bibitem[{{Dekel} {et~al.}(2020){Dekel}, {Lapiner}, {Ginzburg}, {Freundlich},
  {Jiang}, {Finish}, {Kretschmer}, {Lin}, {Ceverino}, {Primack}, {Giavalisco},
  \& {Ji}}]{Dekel2020}
{Dekel}, A., {Lapiner}, S., {Ginzburg}, O., {et~al.} 2020, \mnras, 496, 5372,
  \dodoi{10.1093/mnras/staa1713}

\bibitem[{{Dimauro} {et~al.}(2018){Dimauro}, {Huertas-Company}, {Daddi},
  {P{\'e}rez-Gonz{\'a}lez}, {Bernardi}, {Barro}, {Buitrago}, {Caro},
  {Cattaneo}, {Dominguez-S{\'a}nchez}, {Faber}, {H{\"a}u{\ss}ler}, {Kocevski},
  {Koekemoer}, {Koo}, {Lee}, {Mei}, {Margalef-Bentabol}, {Primack},
  {Rodriguez-Puebla}, {Salvato}, {Shankar}, \& {Tuccillo}}]{Dimauro2018}
{Dimauro}, P., {Huertas-Company}, M., {Daddi}, E., {et~al.} 2018, \mnras, 478,
  5410, \dodoi{10.1093/mnras/sty1379}

\bibitem[{{Dom{\'\i}nguez-Palmero} \& {Balcells}(2008)}]{DominguezPalmero2008}
{Dom{\'\i}nguez-Palmero}, L., \& {Balcells}, M. 2008, \aap, 489, 1003,
  \dodoi{10.1051/0004-6361:200809407}

\bibitem[{{Dom{\'\i}nguez-Palmero} \& {Balcells}(2009)}]{DominguezPalmero2009}
---. 2009, \apjl, 694, L69, \dodoi{10.1088/0004-637X/694/1/L69}

\bibitem[{{Dom{\'\i}nguez S{\'a}nchez} {et~al.}(2016){Dom{\'\i}nguez
  S{\'a}nchez}, {P{\'e}rez-Gonz{\'a}lez}, {Esquej}, {Eliche-Moral}, {Barro},
  {Cava}, {Koekemoer}, {Alcalde Pampliega}, {Alonso Herrero}, {Bruzual},
  {Cardiel}, {Cenarro}, {Ceverino}, {Charlot}, \& {Hern{\'a}n
  Caballero}}]{DominguezSanchez2016}
{Dom{\'\i}nguez S{\'a}nchez}, H., {P{\'e}rez-Gonz{\'a}lez}, P.~G., {Esquej},
  P., {et~al.} 2016, \mnras, 457, 3743, \dodoi{10.1093/mnras/stw201}

\bibitem[{{Dubois} {et~al.}(2021){Dubois}, {Beckmann}, {Bournaud}, {Choi},
  {Devriendt}, {Jackson}, {Kaviraj}, {Kimm}, {Kraljic}, {Laigle}, {Martin},
  {Park}, {Peirani}, {Pichon}, {Volonteri}, \& {Yi}}]{Dubois2021}
{Dubois}, Y., {Beckmann}, R., {Bournaud}, F., {et~al.} 2021, \aap, 651, A109,
  \dodoi{10.1051/0004-6361/202039429}

\bibitem[{{Eggen} {et~al.}(1962){Eggen}, {Lynden-Bell}, \&
  {Sandage}}]{Eggen1962}
{Eggen}, O.~J., {Lynden-Bell}, D., \& {Sandage}, A.~R. 1962, \apj, 136, 748,
  \dodoi{10.1086/147433}

\bibitem[{{El-Badry} {et~al.}(2018){El-Badry}, {Quataert}, {Wetzel}, {Hopkins},
  {Weisz}, {Chan}, {Fitts}, {Boylan-Kolchin}, {Kere{\v{s}}},
  {Faucher-Gigu{\`e}re}, \& {Garrison-Kimmel}}]{ElBadry2018}
{El-Badry}, K., {Quataert}, E., {Wetzel}, A., {et~al.} 2018, \mnras, 473, 1930,
  \dodoi{10.1093/mnras/stx2482}

\bibitem[{{Estrada-Carpenter} {et~al.}(2019){Estrada-Carpenter}, {Papovich},
  {Momcheva}, {Brammer}, {Long}, {Quadri}, {Bridge}, {Dickinson}, {Ferguson},
  {Finkelstein}, {Giavalisco}, {Gosmeyer}, {Lotz}, {Salmon}, {Skelton},
  {Trump}, \& {Weiner}}]{EstradaCarpenter2019}
{Estrada-Carpenter}, V., {Papovich}, C., {Momcheva}, I., {et~al.} 2019, \apj,
  870, 133, \dodoi{10.3847/1538-4357/aaf22e}

\bibitem[{{Estrada-Carpenter} {et~al.}(2020){Estrada-Carpenter}, {Papovich},
  {Momcheva}, {Brammer}, {Simons}, {Bridge}, {Cleri}, {Ferguson},
  {Finkelstein}, {Giavalisco}, {Jung}, {Matharu}, {Trump}, \&
  {Weiner}}]{EstradaCarpenter2020}
---. 2020, \apj, 898, 171, \dodoi{10.3847/1538-4357/aba004}

\bibitem[{{Fall} \& {Efstathiou}(1980)}]{Fall1980}
{Fall}, S.~M., \& {Efstathiou}, G. 1980, \mnras, 193, 189,
  \dodoi{10.1093/mnras/193.2.189}

\bibitem[{{Freeman}(1970)}]{Freeman1970}
{Freeman}, K.~C. 1970, \apj, 160, 811

\bibitem[{{Gadotti} {et~al.}(2020){Gadotti}, {Bittner}, {Falc{\'o}n-Barroso},
  {M{\'e}ndez-Abreu}, {Kim}, {Fragkoudi}, {de Lorenzo-C{\'a}ceres}, {Leaman},
  {Neumann}, {Querejeta}, {S{\'a}nchez-Bl{\'a}zquez}, {Martig},
  {Mart{\'\i}n-Navarro}, {P{\'e}rez}, {Seidel}, \& {van de Ven}}]{Gadotti2020}
{Gadotti}, D.~A., {Bittner}, A., {Falc{\'o}n-Barroso}, J., {et~al.} 2020, \aap,
  643, A14, \dodoi{10.1051/0004-6361/202038448}

\bibitem[{{Gao} {et~al.}(2020){Gao}, {Ho}, {Barth}, \& {Li}}]{Gao2020}
{Gao}, H., {Ho}, L.~C., {Barth}, A.~J., \& {Li}, Z.-Y. 2020, \apjs, 247, 20,
  \dodoi{10.3847/1538-4365/ab67b2}

\bibitem[{Graham(2013)}]{Graham2013}
Graham, A.~W. 2013, Planets, Stars and Stellar Systems, 91–139,
  \dodoi{10.1007/978-94-007-5609-0_2}

\bibitem[{{Graham} \& {Driver}(2005)}]{Graham2005}
{Graham}, A.~W., \& {Driver}, S.~P. 2005, \pasa, 22, 118,
  \dodoi{10.1071/AS05001}

\bibitem[{{Grogin} {et~al.}(2011){Grogin}, {Kocevski}, {Faber}, {Ferguson},
  {Koekemoer}, {Riess}, {Acquaviva}, {Alexander}, {Almaini}, {Ashby}, {Barden},
  {Bell}, {Bournaud}, {Brown}, {Caputi}, {Casertano}, {Cassata}, {Castellano},
  {Challis}, {Chary}, {Cheung}, {Cirasuolo}, {Conselice}, {Roshan Cooray},
  {Croton}, {Daddi}, {Dahlen}, {Dav{\'e}}, {de Mello}, {Dekel}, {Dickinson},
  {Dolch}, {Donley}, {Dunlop}, {Dutton}, {Elbaz}, {Fazio}, {Filippenko},
  {Finkelstein}, {Fontana}, {Gardner}, {Garnavich}, {Gawiser}, {Giavalisco},
  {Grazian}, {Guo}, {Hathi}, {H{\"a}ussler}, {Hopkins}, {Huang}, {Huang},
  {Jha}, {Kartaltepe}, {Kirshner}, {Koo}, {Lai}, {Lee}, {Li}, {Lotz}, {Lucas},
  {Madau}, {McCarthy}, {McGrath}, {McIntosh}, {McLure}, {Mobasher},
  {Moustakas}, {Mozena}, {Nandra}, {Newman}, {Niemi}, {Noeske}, {Papovich},
  {Pentericci}, {Pope}, {Primack}, {Rajan}, {Ravindranath}, {Reddy}, {Renzini},
  {Rix}, {Robaina}, {Rodney}, {Rosario}, {Rosati}, {Salimbeni}, {Scarlata},
  {Siana}, {Simard}, {Smidt}, {Somerville}, {Spinrad}, {Straughn}, {Strolger},
  {Telford}, {Teplitz}, {Trump}, {van der Wel}, {Villforth}, {Wechsler},
  {Weiner}, {Wiklind}, {Wild}, {Wilson}, {Wuyts}, {Yan}, \& {Yun}}]{Grogin2011}
{Grogin}, N.~A., {Kocevski}, D.~D., {Faber}, S.~M., {et~al.} 2011, \apjs, 197,
  35, \dodoi{10.1088/0067-0049/197/2/35}

\bibitem[{{Heavens} {et~al.}(2004){Heavens}, {Panter}, {Jimenez}, \&
  {Dunlop}}]{Heavens2004}
{Heavens}, A., {Panter}, B., {Jimenez}, R., \& {Dunlop}, J. 2004, \nat, 428,
  625, \dodoi{10.1038/nature02474}

\bibitem[{{Hopkins} {et~al.}(2009{\natexlab{a}}){Hopkins}, {Cox}, {Younger}, \&
  {Hernquist}}]{Hopkins2009b}
{Hopkins}, P.~F., {Cox}, T.~J., {Younger}, J.~D., \& {Hernquist}, L.
  2009{\natexlab{a}}, \apj, 691, 1168, \dodoi{10.1088/0004-637X/691/2/1168}

\bibitem[{{Hopkins} {et~al.}(2014){Hopkins}, {Kere{\v{s}}}, {O{\~n}orbe},
  {Faucher-Gigu{\`e}re}, {Quataert}, {Murray}, \& {Bullock}}]{Hopkins2014}
{Hopkins}, P.~F., {Kere{\v{s}}}, D., {O{\~n}orbe}, J., {et~al.} 2014, \mnras,
  445, 581, \dodoi{10.1093/mnras/stu1738}

\bibitem[{{Hopkins} {et~al.}(2009{\natexlab{b}}){Hopkins}, {Somerville}, {Cox},
  {Hernquist}, {Jogee}, {Kere{\v{s}}}, {Ma}, {Robertson}, \&
  {Stewart}}]{Hopkins2009a}
{Hopkins}, P.~F., {Somerville}, R.~S., {Cox}, T.~J., {et~al.}
  2009{\natexlab{b}}, \mnras, 397, 802,
  \dodoi{10.1111/j.1365-2966.2009.14983.x}

\bibitem[{{Hsu} {et~al.}(2019){Hsu}, {Lin}, {Dickinson}, {Yan}, {Bau-Ching},
  {Wang}, {Lee}, {Yan}, {Scott}, {Willner}, {Ouchi}, {Ashby}, {Chen}, {Daddi},
  {Elbaz}, {Fazio}, {Foucaud}, {Huang}, {Koo}, {Morrison}, {Owen}, {Pannella},
  {Pope}, {Simard}, \& {Wang}}]{Hsu2019}
{Hsu}, L.-T., {Lin}, L., {Dickinson}, M., {et~al.} 2019, \apj, 871, 233,
  \dodoi{10.3847/1538-4357/aaf9a7}

\bibitem[{{Huang} {et~al.}(2013{\natexlab{a}}){Huang}, {Ho}, {Peng}, {Li}, \&
  {Barth}}]{Huang2013a}
{Huang}, S., {Ho}, L.~C., {Peng}, C.~Y., {Li}, Z.-Y., \& {Barth}, A.~J.
  2013{\natexlab{a}}, \apj, 766, 47, \dodoi{10.1088/0004-637X/766/1/47}

\bibitem[{{Huang} {et~al.}(2013{\natexlab{b}}){Huang}, {Ho}, {Peng}, {Li}, \&
  {Barth}}]{Huang2013b}
---. 2013{\natexlab{b}}, \apjl, 768, L28, \dodoi{10.1088/2041-8205/768/2/L28}

\bibitem[{{Hubble}(1926)}]{Hubble1926}
{Hubble}, E.~P. 1926, \apj, 64, \dodoi{10.1086/143018}

\bibitem[{{Koekemoer} {et~al.}(2011){Koekemoer}, {Faber}, {Ferguson}, {Grogin},
  {Kocevski}, {Koo}, {Lai}, {Lotz}, {Lucas}, {McGrath}, {Ogaz}, {Rajan},
  {Riess}, {Rodney}, {Strolger}, {Casertano}, {Castellano}, {Dahlen},
  {Dickinson}, {Dolch}, {Fontana}, {Giavalisco}, {Grazian}, {Guo}, {Hathi},
  {Huang}, {van der Wel}, {Yan}, {Acquaviva}, {Alexander}, {Almaini}, {Ashby},
  {Barden}, {Bell}, {Bournaud}, {Brown}, {Caputi}, {Cassata}, {Challis},
  {Chary}, {Cheung}, {Cirasuolo}, {Conselice}, {Roshan Cooray}, {Croton},
  {Daddi}, {Dav{\'e}}, {de Mello}, {de Ravel}, {Dekel}, {Donley}, {Dunlop},
  {Dutton}, {Elbaz}, {Fazio}, {Filippenko}, {Finkelstein}, {Frazer}, {Gardner},
  {Garnavich}, {Gawiser}, {Gruetzbauch}, {Hartley}, {H{\"a}ussler},
  {Herrington}, {Hopkins}, {Huang}, {Jha}, {Johnson}, {Kartaltepe},
  {Khostovan}, {Kirshner}, {Lani}, {Lee}, {Li}, {Madau}, {McCarthy},
  {McIntosh}, {McLure}, {McPartland}, {Mobasher}, {Moreira}, {Mortlock},
  {Moustakas}, {Mozena}, {Nandra}, {Newman}, {Nielsen}, {Niemi}, {Noeske},
  {Papovich}, {Pentericci}, {Pope}, {Primack}, {Ravindranath}, {Reddy},
  {Renzini}, {Rix}, {Robaina}, {Rosario}, {Rosati}, {Salimbeni}, {Scarlata},
  {Siana}, {Simard}, {Smidt}, {Snyder}, {Somerville}, {Spinrad}, {Straughn},
  {Telford}, {Teplitz}, {Trump}, {Vargas}, {Villforth}, {Wagner}, {Wand ro},
  {Wechsler}, {Weiner}, {Wiklind}, {Wild}, {Wilson}, {Wuyts}, \&
  {Yun}}]{Koekemoer2011}
{Koekemoer}, A.~M., {Faber}, S.~M., {Ferguson}, H.~C., {et~al.} 2011, \apjs,
  197, 36, \dodoi{10.1088/0067-0049/197/2/36}

\bibitem[{{Kormendy}(2016)}]{Kormendy2016}
{Kormendy}, J. 2016, Astrophysics and Space Science Library, Vol. 418,
  {Elliptical Galaxies and Bulges of Disc Galaxies: Summary of Progress and
  Outstanding Issues}, ed. E.~{Laurikainen}, R.~{Peletier}, \& D.~{Gadotti},
  431, \dodoi{10.1007/978-3-319-19378-6_16}

\bibitem[{{Kormendy} \& {Kennicutt}(2004)}]{Kormendy2004}
{Kormendy}, J., \& {Kennicutt}, Robert~C., J. 2004, \araa, 42, 603,
  \dodoi{10.1146/annurev.astro.42.053102.134024}

\bibitem[{{Larson}(1976)}]{Larson1976}
{Larson}, R.~B. 1976, \mnras, 176, 31, \dodoi{10.1093/mnras/176.1.31}

\bibitem[{{Lilly} {et~al.}(1996){Lilly}, {Le Fevre}, {Hammer}, \&
  {Crampton}}]{Lilly1996}
{Lilly}, S.~J., {Le Fevre}, O., {Hammer}, F., \& {Crampton}, D. 1996, \apjl,
  460, L1, \dodoi{10.1086/309975}

\bibitem[{{Madau} \& {Dickinson}(2014)}]{Madau2014}
{Madau}, P., \& {Dickinson}, M. 2014, \araa, 52, 415,
  \dodoi{10.1146/annurev-astro-081811-125615}

\bibitem[{{Madau} {et~al.}(1996){Madau}, {Ferguson}, {Dickinson}, {Giavalisco},
  {Steidel}, \& {Fruchter}}]{Madau1996}
{Madau}, P., {Ferguson}, H.~C., {Dickinson}, M.~E., {et~al.} 1996, \mnras, 283,
  1388, \dodoi{10.1093/mnras/283.4.1388}

\bibitem[{{Mancini} {et~al.}(2019){Mancini}, {Daddi}, {Juneau}, {Renzini},
  {Rodighiero}, {Cappellari}, {Rodr{\'\i}guez-Mu{\~n}oz}, {Liu}, {Pannella},
  {Baronchelli}, {Franceschini}, {Bergamini}, {D'Eugenio}, \&
  {Puglisi}}]{Mancini2019}
{Mancini}, C., {Daddi}, E., {Juneau}, S., {et~al.} 2019, \mnras, 489, 1265,
  \dodoi{10.1093/mnras/stz2130}

\bibitem[{{Margalef-Bentabol} {et~al.}(2016){Margalef-Bentabol}, {Conselice},
  {Mortlock}, {Hartley}, {Duncan}, {Ferguson}, {Dekel}, \&
  {Primack}}]{MargalefBentabol2016}
{Margalef-Bentabol}, B., {Conselice}, C.~J., {Mortlock}, A., {et~al.} 2016,
  \mnras, 461, 2728, \dodoi{10.1093/mnras/stw1451}

\bibitem[{{Margalef-Bentabol} {et~al.}(2018){Margalef-Bentabol}, {Conselice},
  {Mortlock}, {Hartley}, {Duncan}, {Kennedy}, {Kocevski}, \&
  {Hasinger}}]{MargalefBentabol2018}
---. 2018, \mnras, 473, 5370, \dodoi{10.1093/mnras/stx2633}

\bibitem[{{M{\'e}ndez-Abreu} {et~al.}(2008){M{\'e}ndez-Abreu}, {Aguerri},
  {Corsini}, \& {Simonneau}}]{MendezAbreu2008}
{M{\'e}ndez-Abreu}, J., {Aguerri}, J.~A.~L., {Corsini}, E.~M., \& {Simonneau},
  E. 2008, \aap, 478, 353, \dodoi{10.1051/0004-6361:20078089}

\bibitem[{{M{\'e}ndez-Abreu} {et~al.}(2021){M{\'e}ndez-Abreu}, {de
  Lorenzo-C{\'a}ceres}, \& {S{\'a}nchez}}]{MendezAbreu2021}
{M{\'e}ndez-Abreu}, J., {de Lorenzo-C{\'a}ceres}, A., \& {S{\'a}nchez}, S.~F.
  2021, \mnras, 504, 3058, \dodoi{10.1093/mnras/stab1064}

\bibitem[{{M{\'e}ndez-Abreu} {et~al.}(2014){M{\'e}ndez-Abreu}, {Debattista},
  {Corsini}, \& {Aguerri}}]{MendezAbreu2014}
{M{\'e}ndez-Abreu}, J., {Debattista}, V.~P., {Corsini}, E.~M., \& {Aguerri},
  J.~A.~L. 2014, \aap, 572, A25, \dodoi{10.1051/0004-6361/201423955}

\bibitem[{{M{\'e}ndez-Abreu} {et~al.}(2019{\natexlab{a}}){M{\'e}ndez-Abreu},
  {S{\'a}nchez}, \& {de Lorenzo-C{\'a}ceres}}]{MendezAbreu2019a}
{M{\'e}ndez-Abreu}, J., {S{\'a}nchez}, S.~F., \& {de Lorenzo-C{\'a}ceres}, A.
  2019{\natexlab{a}}, \mnras, 484, 4298, \dodoi{10.1093/mnras/stz276}

\bibitem[{{M{\'e}ndez-Abreu} {et~al.}(2019{\natexlab{b}}){M{\'e}ndez-Abreu},
  {S{\'a}nchez}, \& {de Lorenzo-C{\'a}ceres}}]{MendezAbreu2019b}
---. 2019{\natexlab{b}}, \mnras, 488, L80, \dodoi{10.1093/mnrasl/slz103}

\bibitem[{{M{\'e}ndez-Abreu} {et~al.}(2010){M{\'e}ndez-Abreu}, {Simonneau},
  {Aguerri}, \& {Corsini}}]{MendezAbreu2010}
{M{\'e}ndez-Abreu}, J., {Simonneau}, E., {Aguerri}, J.~A.~L., \& {Corsini},
  E.~M. 2010, \aap, 521, A71, \dodoi{10.1051/0004-6361/201014130}

\bibitem[{{Mo} {et~al.}(1998){Mo}, {Mao}, \& {White}}]{Mo1998}
{Mo}, H.~J., {Mao}, S., \& {White}, S. D.~M. 1998, \mnras, 295, 319,
  \dodoi{10.1046/j.1365-8711.1998.01227.x}

\bibitem[{{Morelli} {et~al.}(2015){Morelli}, {Corsini}, {Pizzella}, {Dalla
  Bont{\`a}}, {Coccato}, \& {M{\'e}ndez-Abreu}}]{Morelli2015}
{Morelli}, L., {Corsini}, E.~M., {Pizzella}, A., {et~al.} 2015, \mnras, 452,
  1128, \dodoi{10.1093/mnras/stv1357}

\bibitem[{{Morelli} {et~al.}(2016){Morelli}, {Parmiggiani}, {Corsini},
  {Costantin}, {Dalla Bont{\`a}}, {M{\'e}ndez-Abreu}, \&
  {Pizzella}}]{Morelli2016}
{Morelli}, L., {Parmiggiani}, M., {Corsini}, E.~M., {et~al.} 2016, \mnras, 463,
  4396, \dodoi{10.1093/mnras/stw2285}

\bibitem[{{Morishita} {et~al.}(2019){Morishita}, {Abramson}, {Treu}, {Brammer},
  {Jones}, {Kelly}, {Stiavelli}, {Trenti}, {Vulcani}, \&
  {Wang}}]{Morishita2019}
{Morishita}, T., {Abramson}, L.~E., {Treu}, T., {et~al.} 2019, \apj, 877, 141,
  \dodoi{10.3847/1538-4357/ab1d53}

\bibitem[{{Naab} {et~al.}(2009){Naab}, {Johansson}, \& {Ostriker}}]{Naab2009}
{Naab}, T., {Johansson}, P.~H., \& {Ostriker}, J.~P. 2009, \apjl, 699, L178,
  \dodoi{10.1088/0004-637X/699/2/L178}

\bibitem[{{Nelson} {et~al.}(2018){Nelson}, {Pillepich}, {Springel},
  {Weinberger}, {Hernquist}, {Pakmor}, {Genel}, {Torrey}, {Vogelsberger},
  {Kauffmann}, {Marinacci}, \& {Naiman}}]{Nelson2018}
{Nelson}, D., {Pillepich}, A., {Springel}, V., {et~al.} 2018, \mnras, 475, 624,
  \dodoi{10.1093/mnras/stx3040}

\bibitem[{{Noguchi}(1999)}]{Noguchi1999}
{Noguchi}, M. 1999, \apj, 514, 77, \dodoi{10.1086/306932}

\bibitem[{{Oser} {et~al.}(2010){Oser}, {Ostriker}, {Naab}, {Johansson}, \&
  {Burkert}}]{Oser2010}
{Oser}, L., {Ostriker}, J.~P., {Naab}, T., {Johansson}, P.~H., \& {Burkert}, A.
  2010, \apj, 725, 2312, \dodoi{10.1088/0004-637X/725/2/2312}

\bibitem[{{Park} {et~al.}(2019){Park}, {Yi}, {Dubois}, {Pichon}, {Kimm},
  {Devriendt}, {Choi}, {Volonteri}, {Kaviraj}, \& {Peirani}}]{Park2019}
{Park}, M.-J., {Yi}, S.~K., {Dubois}, Y., {et~al.} 2019, \apj, 883, 25,
  \dodoi{10.3847/1538-4357/ab3afe}

\bibitem[{{Patel} {et~al.}(2013){Patel}, {van Dokkum}, {Franx}, {Quadri},
  {Muzzin}, {Marchesini}, {Williams}, {Holden}, \& {Stefanon}}]{Patel2013}
{Patel}, S.~G., {van Dokkum}, P.~G., {Franx}, M., {et~al.} 2013, \apj, 766, 15,
  \dodoi{10.1088/0004-637X/766/1/15}

\bibitem[{{P{\'e}rez-Gonz{\'a}lez} {et~al.}(2003){P{\'e}rez-Gonz{\'a}lez}, {Gil
  de Paz}, {Zamorano}, {Gallego}, {Alonso-Herrero}, \&
  {Arag{\'o}n-Salamanca}}]{PerezGonzalez2003}
{P{\'e}rez-Gonz{\'a}lez}, P.~G., {Gil de Paz}, A., {Zamorano}, J., {et~al.}
  2003, \mnras, 338, 525, \dodoi{10.1046/j.1365-8711.2003.06078.x}

\bibitem[{{P{\'e}rez-Gonz{\'a}lez} {et~al.}(2008){P{\'e}rez-Gonz{\'a}lez},
  {Rieke}, {Villar}, {Barro}, {Blaylock}, {Egami}, {Gallego}, {Gil de Paz},
  {Pascual}, {Zamorano}, \& {Donley}}]{PerezGonzalez2008}
{P{\'e}rez-Gonz{\'a}lez}, P.~G., {Rieke}, G.~H., {Villar}, V., {et~al.} 2008,
  \apj, 675, 234, \dodoi{10.1086/523690}

\bibitem[{{P{\'e}rez-Gonz{\'a}lez} {et~al.}(2013){P{\'e}rez-Gonz{\'a}lez},
  {Cava}, {Barro}, {Villar}, {Cardiel}, {Ferreras}, {Rodr{\'\i}guez-Espinosa},
  {Alonso-Herrero}, {Balcells}, {Cenarro}, {Cepa}, {Charlot}, {Cimatti},
  {Conselice}, {Daddi}, {Donley}, {Elbaz}, {Espino}, {Gallego}, {Gobat},
  {Gonz{\'a}lez-Mart{\'\i}n}, {Guzm{\'a}n}, {Hern{\'a}n-Caballero},
  {Mu{\~n}oz-Tu{\~n}{\'o}n}, {Renzini}, {Rodr{\'\i}guez-Zaur{\'\i}n}, {Tresse},
  {Trujillo}, \& {Zamorano}}]{PerezGonzalez2013}
{P{\'e}rez-Gonz{\'a}lez}, P.~G., {Cava}, A., {Barro}, G., {et~al.} 2013, \apj,
  762, 46, \dodoi{10.1088/0004-637X/762/1/46}

\bibitem[{{Pfeffer} {et~al.}(2022){Pfeffer}, {Bekki}, {Couch}, {Koribalski}, \&
  {Forbes}}]{Pfeffer2022}
{Pfeffer}, J., {Bekki}, K., {Couch}, W.~J., {Koribalski}, B.~S., \& {Forbes},
  D.~A. 2022, arXiv e-prints, arXiv:2201.03137

\bibitem[{{Pillepich} {et~al.}(2018){Pillepich}, {Nelson}, {Hernquist},
  {Springel}, {Pakmor}, {Torrey}, {Weinberger}, {Genel}, {Naiman}, {Marinacci},
  \& {Vogelsberger}}]{Pillepich2018}
{Pillepich}, A., {Nelson}, D., {Hernquist}, L., {et~al.} 2018, \mnras, 475,
  648, \dodoi{10.1093/mnras/stx3112}

\bibitem[{{S{\'a}nchez} {et~al.}(2012){S{\'a}nchez}, {Kennicutt}, {Gil de Paz},
  {van de Ven}, {V{\'\i}lchez}, {Wisotzki}, {Walcher}, {Mast}, {Aguerri},
  {Albiol-P{\'e}rez}, {Alonso-Herrero}, {Alves}, {Bakos}, {Bart{\'a}kov{\'a}},
  {Bland-Hawthorn}, {Boselli}, {Bomans}, {Castillo-Morales}, {Cortijo-Ferrero},
  {de Lorenzo-C{\'a}ceres}, {Del Olmo}, {Dettmar}, {D{\'\i}az}, {Ellis},
  {Falc{\'o}n-Barroso}, {Flores}, {Gallazzi}, {Garc{\'\i}a-Lorenzo},
  {Gonz{\'a}lez Delgado}, {Gruel}, {Haines}, {Hao}, {Husemann},
  {Igl{\'e}sias-P{\'a}ramo}, {Jahnke}, {Johnson}, {Jungwiert}, {Kalinova},
  {Kehrig}, {Kupko}, {L{\'o}pez-S{\'a}nchez}, {Lyubenova}, {Marino},
  {M{\'a}rmol-Queralt{\'o}}, {M{\'a}rquez}, {Masegosa}, {Meidt},
  {Mendez-Abreu}, {Monreal-Ibero}, {Montijo}, {Mour{\~a}o}, {Palacios-Navarro},
  {Papaderos}, {Pasquali}, {Peletier}, {P{\'e}rez}, {P{\'e}rez}, {Quirrenbach},
  {Rela{\~n}o}, {Rosales-Ortega}, {Roth}, {Ruiz-Lara},
  {S{\'a}nchez-Bl{\'a}zquez}, {Sengupta}, {Singh}, {Stanishev}, {Trager},
  {Vazdekis}, {Viironen}, {Wild}, {Zibetti}, \& {Ziegler}}]{Sanchez2012}
{S{\'a}nchez}, S.~F., {Kennicutt}, R.~C., {Gil de Paz}, A., {et~al.} 2012,
  \aap, 538, A8, \dodoi{10.1051/0004-6361/201117353}

\bibitem[{{Schaye} {et~al.}(2015){Schaye}, {Crain}, {Bower}, {Furlong},
  {Schaller}, {Theuns}, {Dalla Vecchia}, {Frenk}, {McCarthy}, {Helly},
  {Jenkins}, {Rosas-Guevara}, {White}, {Baes}, {Booth}, {Camps}, {Navarro},
  {Qu}, {Rahmati}, {Sawala}, {Thomas}, \& {Trayford}}]{Schaye2015}
{Schaye}, J., {Crain}, R.~A., {Bower}, R.~G., {et~al.} 2015, \mnras, 446, 521,
  \dodoi{10.1093/mnras/stu2058}

\bibitem[{{S{\'e}rsic}(1968)}]{Sersic1968}
{S{\'e}rsic}, J.~L. 1968, {Atlas de Galaxias Australes} (Observatorio
  Astronomico de Cordoba, Cordoba)

\bibitem[{{Steinmetz} \& {Navarro}(2002)}]{Steinmetz2002}
{Steinmetz}, M., \& {Navarro}, J.~F. 2002, \na, 7, 155,
  \dodoi{10.1016/S1384-1076(02)00102-1}

\bibitem[{{Suess} {et~al.}(2021){Suess}, {Kriek}, {Price}, \&
  {Barro}}]{Suess2021}
{Suess}, K.~A., {Kriek}, M., {Price}, S.~H., \& {Barro}, G. 2021, \apj, 915,
  87, \dodoi{10.3847/1538-4357/abf1e4}

\bibitem[{{Tacchella} {et~al.}(2019){Tacchella}, {Diemer}, {Hernquist},
  {Genel}, {Marinacci}, {Nelson}, {Pillepich}, {Rodriguez-Gomez}, {Sales},
  {Springel}, \& {Vogelsberger}}]{Tacchella2019}
{Tacchella}, S., {Diemer}, B., {Hernquist}, L., {et~al.} 2019, \mnras, 487,
  5416, \dodoi{10.1093/mnras/stz1657}

\bibitem[{{Tacchella} {et~al.}(2021){Tacchella}, {Conroy}, {Faber}, {Johnson},
  {Leja}, {Barro}, {Cunningham}, {Deason}, {Guhathakurta}, {Guo}, {Hernquist},
  {Koo}, {McKinnon}, {Rockosi}, {Speagle}, {van Dokkum}, \&
  {Yesuf}}]{Tacchella2021}
{Tacchella}, S., {Conroy}, C., {Faber}, S.~M., {et~al.} 2021, arXiv e-prints,
  arXiv:2102.12494

\bibitem[{{van Dokkum} {et~al.}(2010){van Dokkum}, {Whitaker}, {Brammer},
  {Franx}, {Kriek}, {Labb{\'e}}, {Marchesini}, {Quadri}, {Bezanson},
  {Illingworth}, {Muzzin}, {Rudnick}, {Tal}, \& {Wake}}]{vanDokkum2010}
{van Dokkum}, P.~G., {Whitaker}, K.~E., {Brammer}, G., {et~al.} 2010, \apj,
  709, 1018, \dodoi{10.1088/0004-637X/709/2/1018}

\bibitem[{{Williams} {et~al.}(2010){Williams}, {Quadri}, {Franx}, {van Dokkum},
  {Toft}, {Kriek}, \& {Labb{\'e}}}]{Williams2010}
{Williams}, R.~J., {Quadri}, R.~F., {Franx}, M., {et~al.} 2010, \apj, 713, 738,
  \dodoi{10.1088/0004-637X/713/2/738}

\bibitem[{{Zolotov} {et~al.}(2015){Zolotov}, {Dekel}, {Mandelker}, {Tweed},
  {Inoue}, {DeGraf}, {Ceverino}, {Primack}, {Barro}, \& {Faber}}]{Zolotov2015}
{Zolotov}, A., {Dekel}, A., {Mandelker}, N., {et~al.} 2015, \mnras, 450, 2327,
  \dodoi{10.1093/mnras/stv740}

\end{thebibliography}
\end{document}